\documentclass[a4paper,UKenglish,cleveref, autoref, thm-restate]{lipics-v2021}

\usepackage{sections/camera-ready-macros}

\hideLIPIcs  

\title{Individual Fairness in Advertising Auctions through Inverse Proportionality} 

\author{Shuchi Chawla}{The University of Texas at Austin, USA }{shuchi@cs.utexas.edu}{}{Supported in part by NSF awards CCF-2008006 and SHF-1704117.}

\author{Meena Jagadeesan}{University of California, Berkeley, USA }{mjagadeesan@berkeley.edu}{}{Supported by a Paul and Daisy Soros Fellowship.}

\authorrunning{Shuchi Chawla and Meena Jagadeesan} 

\Copyright{Shuchi Chawla and Meena Jagadeesan} 

\ccsdesc[500]{Theory of computation~Algorithmic mechanism design}

\keywords{Algorithmic fairness, advertising auctions} 

\category{} 

\nolinenumbers 

\begin{document}

\maketitle

\begin{abstract}
Recent empirical work demonstrates that online advertisement can exhibit bias in the delivery of ads across users even when all advertisers bid in a non-discriminatory manner. We study the design of ad auctions that, given fair bids, are guaranteed to produce fair outcomes. Following the works of Dwork and Ilvento \cite{DI2018} and Chawla et al. \cite{CIJ20}, our goal is to design a truthful auction that satisfies ``individual fairness'' in its outcomes: informally speaking, users that are similar to each other should obtain similar allocations of ads. Within this framework we quantify the tradeoff between social welfare maximization and fairness.

This work makes two conceptual contributions. First, we express the fairness constraint as a kind of stability condition: any two users that are assigned multiplicatively similar values by all the advertisers must receive additively similar allocations for each advertiser. This value stability constraint is expressed as a function that maps the multiplicative distance between value vectors to the maximum allowable $\ell_{\infty}$ distance between the corresponding allocations. Standard auctions do not satisfy this kind of value stability.

Second, we introduce a new class of allocation algorithms called Inverse Proportional Allocation that achieve a near optimal tradeoff between fairness and social welfare for a broad and expressive class of value stability conditions. These allocation algorithms are truthful and prior-free, and achieve a constant factor approximation to the optimal (unconstrained) social welfare. In particular, the approximation ratio is independent of the number of advertisers in the system. In this respect, these allocation algorithms greatly surpass the guarantees achieved in previous work. We also extend our results to broader notions of fairness that we call subset fairness.
\end{abstract}

\section{Introduction}\label{sec:introduction}

Algorithms play an increasingly important role in today's society and can arguably have far-ranging social, economic, and political ramifications in many different contexts. One such context is access to sponsored information such as advertisements on social media, search pages, and other websites. It has been well documented that online advertisement can exhibit skewed delivery: users that differ on sensitive attributes such as race, gender, age, religion, and national origin can receive very different allocations of ads (see, e.g., \cite{S13, DTD15, AP2016, AST2017,  LT2016, propub, ASBKMR2019}). For example, certain employment jobs on Facebook have been found to exclusively target men or exclude older people; and housing ads have been found to exclude users based on race~\cite{AST2017, AP2016, propub}.

There are two main sources of unfairness in digital ads. The first is explicit or implicit targeting of users based on sensitive attributes by the advertisers. This source of unfairness can be mitigated by examining and auditing each advertiser's targeting approach separately.\footnote{ Indeed, in response to a lawsuit brought on by the U.S. Department of
Housing and Urban Development and ProPublica against Facebook on discrimination in housing ads, Facebook's first action was to disallow advertisers to target users based on sensitive attributes~\cite{propub}. However, despite this action, unfairness in outcomes continued to persist~\cite{ASBKMR2019}.} The second, more subtle and insidious, source of unfairness is the ad delivery mechanism that determines for each user which ad to display based on advertisers' bids, budgets, relevance of the ads to the user, etc. Recent empirical studies \cite{LT2016, ASBKMR2019} have observed that even when each advertiser's targeting parameters are inclusive, skewed delivery can arise as a result of the ad auction mechanism. Our paper focuses on eliminating this second source of unfairness. 

{\bf We study the design of ad auctions from the perspective of individual fairness}. Individual fairness \cite{DHPRZ2012} posits that similar users should be treated similarly. In the context of sponsored content, this translates into requiring that similar users should be served a similar mix of ads. 
We follow the approach of Chawla et al. \cite{CIJ20} and isolate the effect of {\it auction design} on fairness in outcomes by assuming that each advertiser's bids are non-discriminatory. Then, if two users receive similar bids from all advertisers, we require that they receive similar ad allocations. In other words, the auction does not introduce any further unfairness than what may be present already in advertisers' bids. 

We consider a stylized model of (truthful) ad auctions where users arrive over time and each is shown a single ad. When a user arrives, $k$ advertisers report their values for the single ad slot; denote these values with the vector $\val=(\vali[1], \vali[2], \cdots, \vali[k])$.\footnote{We focus in this work on truthful auctions, and so do not make a distinction between bids and values. All of the allocation rules we propose and study are monotone in values and can be implemented in a truthful manner.} The auction selects a (potentially random) advertiser based on the values and displays its ad; we denote the auction's selection by a distribution over $[k]$, $\alloc(\val)\in \Delta([k])$, where $\alloci(\val)$ is the probability that advertiser $i$'s bid is displayed. We measure the quality of this allocation in terms of its social welfare -- the expected sum of values corresponding to ads displayed, with the contribution of each user to this sum given by $\val\cdot\alloc(\val) = \sum_{i\in [k]} \vali\alloci(\val)$. 

Social welfare maximization for our simple model in the absence of a fairness constraint is easy: just serve each user the ad with the highest value. Real-world ad auctions are essentially variants of this ``highest-bid-wins'' auction, modified to accommodate click-through-rates, multiple slots, etc. Our main challenge is that these standard auction formats are fundamentally incompatible with fairness. Consider, for example, two job ads A and B competing over two users Alice and Bob. Suppose that Alice and Bob have similar educational qualifications and work experience but belong to different ethnic groups. Further, due to slight differences in their profiles, suppose that A values Alice slightly higher than B and B values Bob slightly higher than A. Then a highest-bid-wins auction would allocate ad A to Alice and ad B to Bob. This violates our fairness constraint: although the two users are similar to each other in most respects, their ad allocations are completely different. Indeed, if job B is much more desirable than job A, then Alice is discriminated against. Note that the unfairness persists even if each advertiser individually places similar values on the two users. As we show in Section~\ref{sec:experiments}, this behavior is not just hypothetical but is frequently observed in real data as well. The highest-bid-wins auction greatly exaggerates minor, subtle differences in input into huge swings in output. 

\subsection{Our contributions}
In this paper we propose a new framework for expressing fairness in ad auctions as a value-stability condition. We then design allocation algorithms that achieve near-optimal tradeoffs between value-stability and social welfare. 

\paragraph{Fairness as a stability condition.}  Consider, again, two users Alice and Bob. Let $\val$ denote the values received by Alice and $\val'$ the values received by Bob. We will use a parameter $\stabpar\ge 1$ to quantify the similarity between Alice and Bob, and will say that they receive non-discriminatory values if it holds that $\vali'\in [\frac 1\stabpar \vali, \stabpar\vali]$ for all $i\in [k]$. Given such non-discriminatory values, we will require that Alice and Bob receive allocations that are component-wise close to each other, i.e. $\alloci(\val')\in \alloci(\val)\pm f(\stabpar)$ for all $i\in [k]$.\footnote{We emphasize that we do not require that two ads with similar values for the same user be served with similar probabilities: we aim to satisfy fairness across users, not across advertisers.} Here $f$ is an externally specified function that we call the fairness constraint and that governs the strength of the fairness guarantee. We say that the auction is {\em value stable} with respect to the constraint $f$. 

We essentially require the allocation function to satisfy a kind of stability condition: when it receives two value vectors that are component-wise close, it should return allocations that are component-wise close. Observe that we measure closeness in values as a ratio, and consequently our stability notion is scale free. On the other hand, we measure closeness in allocations as a difference, specifically the $\ell_\infty$ distance between the respective distributions. This corresponds to the notion of {\em multiple-task fairness} defined by Dwork and Ilvento \cite{DI2018} in the context of multi-dimensional allocation algorithms. 


Our approach of expressing fairness as a stability constraint offers several benefits. First, our approach assumes that every pair of similar users receives multiplicatively similar values from {\em all} advertisers. This enables a clear separation of responsibility between the auctioneer and the advertisers, which can simplify the design of both of these components of the ad auction pipeline. We envision that an auditor would check whether advertisers are following fairness guidelines. As a result, the auctioneer only needs to ensure that the allocation is fair when the advertisers' reported values are fair, but is required to provide no fairness guarantee when the values themselves are unfair. Second, and more importantly, the auctioneer can provide this guarantee without access to the specifics of fairness requirements across users. One criticism of the notion of individual fairness is its reliance on an appropriate similarity metric on users that captures exactly what features can be used to differentiate between users. This metric can not only be challenging to learn, but may itself be sensitive information that is unavailable to the auctioneer or illegal to use directly. Our approach allows the auctioneer to provide a meaningful fairness guarantee without knowing or using the similarity metric. In effect, the metric is encoded in the advertisers' reported values when those values are fair.

Finally, we note that in \cite{CIJ20}, we considered a similar model of fairness in ad auctions but formalized the fairness constraint in a slightly different manner. The algorithmic results in this paper are directly comparable to those in \cite{CIJ20}, and we make the connection explicit in Section~\ref{sec:fairness-stability}.



\paragraph{Designing value-stable allocation algorithms.} 
Now that we have defined value-stability, our goal can be restated as designing value stable auctions that achieve high social welfare.

In \cite{CIJ20}, we proposed a natural candidate for value stability, namely Proportional Allocation (PA) algorithm: given the value vector $\val$, we assign each ad $i$ to the user with probability $\vali/\sum_j \vali[j]$. It is straightforward to observe that changing each value by a factor of $\stabpar$ changes each allocation multiplicatively by at most $\stabpar^2$. In fact, PA actually achieves the additive form of fairness defined above with $f(\stabpar)=(\stabpar-1)/(\stabpar+1)$.\footnote{The strength of the fairness guarantee can be changed by allocating each ad with probability proportional to an appropriate function of the corresponding value.} Unfortunately, the social welfare of PA degrades as the number of advertisers increases. Suppose, for example, that there are $k+1$ advertisers; the first one places a value of $1$ on the user whereas all the others value the user at $x<1$. Then, PA assigns the first ad to the user with probability $1/(1+xk)$ and obtains a social welfare of $(1+x^2k)/(1+xk)$. When $x=1/\sqrt{k}$, this is a factor of $\sqrt{k}$ off from the optimal social welfare of $1$. Other proportional allocation algorithms that assign allocations in proportion to other functions of values\footnote{This class includes, for example, the exponential algorithm from differential privacy literature \cite{ST07}.} suffer from the same problem:  their worst case approximation ratio goes to zero as the number $k$ of advertisers goes to infinity. Is it possible to obtain better social welfare while maintaining fairness?

In this paper we introduce a new family of allocation rules that we call {\bf Inverse Proportional Allocation} (IPA). Informally, our auction begins by allocating each ad fully (but infeasibly) to the user. It then ``takes away'' the over-assignment from the ads in proportion to some decreasing function of the advertisers' values until a total allocation probability of $1$ is achieved. The choice of the decreasing function depends on the strength of the desired fairness constraint. This new family of mechanisms achieves fairness guarantees similar to those of PA, while obtaining a  {\em constant factor approximation} to the unfair optimal social welfare, independent of the number of advertisers $k$. We show, in fact, that Inverse Proportional Allocation achieves a near-optimal tradeoff between fairness and social welfare for {\em any} stability constraint satisfying a mild assumption. 

We describe the fairness guarantee of our algorithm in the form of a ``fairness constraint'' $f$ that specifies how different the allocations for two users can be as a function of the similarity between the users. We use a parameter $\ell \in (0, \infty)$ to denote the strength of this fairness constraint: the smaller that $\ell$ is, the stronger is the fairness guarantee. $\ell=0$ corresponds to perfect fairness or the function that is $0$ at all distances: all users receive identical allocations; and $\ell=\infty$ corresponds to unconstrained allocations or the function that is $1$ at all distances: users can receive arbitrary allocations. For intermediate $\ell$, we consider a polynomial family $f_\ell$ of fairness constraints that best captures the fairness profile of IPA (parameterized appropriately).

\begin{theorem}[Informal]
For any $\ell \in (0, \infty)$, Inverse Proportional Allocation with parameter $\ell$ is value stable with respect to $f_\ell(\stabpar) = 1 - \stabpar^{-2\ell}$. Its worst case approximation ratio for social welfare, $\alpha_\ell$, is $3/4$ at $\ell = 1$ and approaches $1$ as $\ell \rightarrow \infty$. No prior free allocation algorithm that is value stable with respect to $f_\ell$ can achieve a better approximation ratio than $\alpha_\ell+1/k$.
\end{theorem}

When we further expand the class of allocation algorithms to include IPA mixed in with the uniform allocation, we achieve near-optimality against a broad class of constraints $f$.
\begin{theorem}[Informal]
Let $f:[1,\infty] \rightarrow [0,1]$ be any non-decreasing function such that $f(x)/\log x$ is non-increasing. Suppose there exists an allocation algorithm that is value stable with respect to $f$ and achieves a worst case approximation factor of $\alpha$. Then there is an IPA algorithm that is value stable with respect to $f$ and achieves a worst case approximation factor of $\Omega(\alpha/(1+\log(1/\alpha)))$.
\end{theorem}

\paragraph{Empirical evaluation.} 
In addition to theoretical analysis, we perform experiments on sponsored search auction data from Yahoo \cite{Yahoo} to corroborate our theoretical bounds with respect to both the social welfare and fairness. One practical challenge for our approach, which relies on values to infer similarity between users, is that not all advertisers bid on all users: some advertisers may run out of budget, and some choose to bid on only a subpopulation. Interpreting missing bids as zero values can cause the algorithm to form an incorrect estimate of the similarity between two users, potentially violating fairness. For our experiments, we define two users as similar as long as there is significant overlap between the sets of advertisers that bid on both (in terms of Jaccard similarity) and the values of those common advertisers are multiplicatively close. 

Our main observations are twofold. On the one hand, the highest-bid-wins auction is far from value stable, exhibiting widely varying allocations even for pairs of users with value ratios very close to $1$. On the other hand, IPA remains fair in the face of missing bids and, as predicted by the theoretical bounds, the degree of value stability weakens as $\ell$ increases. Relative to the optimal social welfare, IPA obtains approximation ratios of over 90\% for a large range of parameter values and up to 98\% in some cases. IPA therefore offers a vast improvement over highest-bid-wins in terms of fairness while suffering a minimal loss in performance.

\paragraph{Comparison between IPA and PA.} While our primary focus is on comparing the IPA with the highest-bid-wins auction, a.k.a. the unfair optimum, our results provide insight into a comparison between IPA and PA as well. In order to obtain an accurate picture, it is important to consider parameterizations of both algorithms that satisfy the same fairness constraint. In our theoretical analysis, we use the fact that for any fixed $\ell$, the fairness constraint $f_\ell$ is satisfied by IPA$_{\ell}$ and PA$_{2\ell}$. When we take a limit as $k \rightarrow \infty$, we show that IPA$_{\ell}$ achieves a constant factor approximation for social welfare, while the worst case social welfare of PA$_{\ell}$ approaches $0$. Moreover, IPA$_{\ell}$ achieves a better worst-case social welfare than PA$_{2\ell}$ as long as $k \ge 6$.  In this sense, IPA greatly surpasses PA in terms of the fairness/social-welfare tradeoffs. 

In our empirical evaluation, we must consider the performance of these algorithms for small constant values of $k$. This is tricky because the theoretical analyses of fairness for both algorithms are loose for smaller values of $k$. To resolve this, we determine a parameterization for PA by empirically matching its fairness profile with that of IPA$_{\ell}$ as closely as possible. For the Yahoo dataset \cite{Yahoo} that we use, a close match occurs between IPA$_{\ell}$ and PA$_{\frac 43\ell}$. We then compare the performance of the two algorithms and find that IPA consistently performs better on social welfare (with the exception of when the number of advertisers in the auction is 2). In fact, as the number of advertisers increases, the performance of IPA remains unchanged, while that of PA degrades rapidly. We conclude that in most settings, IPA provides a better tradeoff between fairness and social welfare.

\subsection{Further extensions}
We also show our results can be extended to the following two settings.

\paragraph{Fairness across different categories of ads.} So far in our discussion we have assumed that there is ansingle notion of similarity across users that all advertisers are required to respect. In reality, platforms often service many different categories of advertisers (for example job advertisers and credit advertisers) that may be subject to different notions of similarity over the users. These advertisers all compete against each other for the same users. Chawla et al. \cite{CIJ20} proposed fairness definitions for the multi-category setting that combine individual fairness with envy-freeness. Our algorithms from this work can be utilized within the multi-category framework in \cite{CIJ20} to achieve significantly improved tradeoffs between fairness and social welfare for the multi-category setting. We discuss this extension in Appendix \ref{appendix:multicategory}.

\paragraph{Total-variation and subset fairness.}  In \cite{CIJ20}, we showed that PA satisfies a stronger fairness guarantee called {\em total variation fairness}: the distributions over ads assigned to similar users are not only close under $\ell_\infty$ distance, but also close under $\ell_1$ or total variation distance. Equivalently, total-variation fairness provides guarantees over any subset of ads, rather than just a single ad. However, we observe that our family of IPA auctions does not always satisfy total variation fairness. Indeed it appears to be challenging to satisfy this stronger fairness property while also guaranteeing a constant factor approximation in social welfare. In Section \ref{sec:subset}, we show, however, that IPA can be adapted to achieve an intermediate fairness guarantee that we call \textit{subset fairness}: fairness is guaranteed over a restricted collection of subsets of advertisers.

\subsection{Other related work}
While algorithmic fairness is by now an established area of study, theoretical work on fairness in ad auctions is relatively new. There are two complementary sides to the fairness problem in the context of ad auctions. As in our work, Celis et al. \cite{celis2019} study the problem from the viewpoint of the platform---redesigning the ad auction so as to achieve fair outcomes. They provide an optimization framework within which, given an explicit population of users and bids over them, a platform can compute the optimal fair assignment of ads to users. Celis et al. focus on algorithmic techniques for solving the optimization problem, comparing the performance of their algorithm to the optimal fair solution; whereas our focus in this work is on comparing the performance of our algorithm against the unfair optimum and therefore characterizing the cost of fairness.

The complementary viewpoint is that of the advertisers---designing bidding and targetting strategies that preemptively correct for unfairness introduced by the platform mechanism. Nasr and Tschantz \cite{NTC2020} design bidding strategies for advertisers aimed at obtaining parity in impressions across fixed categories (such as gender). Gelauff et al. \cite{GGMY20} likewise design targetting strategies aimed at obtaining parity in outcomes or conversions across different demographic groups. A key point of difference between these works and ours is that they focus on the notion of group fairness whereas our work considers individual fairness guarantees.

Besides these works, the tradeoff between fairness and social welfare has also been studied in other contexts such as classification problems (see, e.g., Hu and Chen \cite{HC2020}).

Complementary to algorithmic fairness, there is an extensive literature on fair division that also considers the tension between fairness and social welfare. 
A major difference between the algorithmic fairness and envy-freeness is that algorithmic fairness generally focuses on individual qualifications, while envy-freeness generally focuses on individual preferences, though recent work \cite{CIJ20, KKRY2019, balcan2018envy, zafar2017parity} has proposed definitions which combine aspects of envy-freeness and algorithmic fairness notions. 

\subsection{Outline for the rest of the paper} In Section \ref{sec:model}, we present our model for fair ad auctions and describe the connection between value stability and individual fairness. In Section \ref{sec:inverseprop}, we describe Inverse Proportional Allocation algorithms and state our main results on social welfare and value stability. In Section \ref{sec:invgen}, we prove the value stability results. In Section \ref{sec:genf} we consider extensions of the IPA mixed in with uniform allocation, and show that this family of allocation algorithms achieves near optimal tradeoffs between social welfare and value stability. We present our experimental results in Section~\ref{sec:experiments}. In Section \ref{sec:subset}, we consider the stronger notion of subset fairness and develop fair allocation algorithms with social welfare that degrade with the complexity of the subset collection.

\section{Ad Auctions and Value Stability}\label{sec:model}

We use a simple stylized model for ad auctions. Every time a user arrives on the platform, $k$ advertisers bid for the chance to display their ad in the {\em single} ad slot available. We assume that the platform employs a truthful direct auction to determine which ad to display. In particular, when the user arrives, each advertiser $i$ reports a value $\vali$ for the user. The auction takes the value vector $\val=[\vali[1], \cdots, \vali[k]]$ as input and returns a distribution $\alloc(\val)=[\alloci[1](\val), \cdots, \alloci[k](\val)]$ over the advertisers. We drop the argument $\val$ when it is clear from the context. With probability $\alloci$, advertiser $i$ is picked and the corresponding ad is displayed for the user. We call the function $\alloc(\val)$ the allocation function.

An allocation function $\alloc$ is truthful if there exists an accompanying payment function $\pay(\val)$ that incentivizes advertisers to bid their true values. Such a payment function exists if and only if for all $i$, $\alloci(\vali,\vali[-i])$ is a monotone non-decreasing function of the argument $\vali$; payments can be computed using the standard payment identity. Henceforth we focus on weakly monotone allocation functions. We briefly discuss the computation of payments using the payment identity in Appendix~\ref{sec:payment}.


\paragraph{Value stability.} 
In order to prevent the allocation function from introducing unfairness into the system, we require that our allocation is \textit{value-stable}. If two users have similar values from all advertisers, then they must receive similar allocations. That is, we require that for two value vectors $\val$ and $\val'$ that are component-wise close in a multiplicative sense, the allocations $\alloc(\val)$ and $\alloc(\val')$ must be close in an additive sense. Formally: 
\begin{definition}
\label{def:valuestable}
An allocation is \textbf{value-stable} with respect to function $f: [1, \infty] \rightarrow [0,1]$ if the following condition is satisfied for every pair of value vectors $\val$ and $\val'$:
\[|\alloci(\val) - \alloci(\val')| \le f(\stabpar) \text{ for all } i \in [k], \,\, \text{where }\stabpar \text{ is defined as } \max_{i\in [k]} \left( \max\left\{\frac \vali{\vali'}, \frac {\vali'}\vali\right\}\right).\]
\end{definition}

\noindent
The function $f$ governs the strength of the value stability constraint. When two value vectors are identical, the allocation should be the same, so $f(1) = 0$. When two value vectors are arbitrarily different, the allocation should be permitted to be arbitrarily different, so $f(\infty) = 1$. For intermediate values, observe that the smaller $f$ is, the tighter is the constraint on the allocation, and the harder it should be for the auction to achieve value stability. Following \cite{CIJ20}, we characterize the strength of the fairness constraint on values by considering an explicit family $\mathcal F$ of functions parameterized by $\ell\in (0,\infty)$: $f_\ell(x) = 1 - x^{-2\ell}$.\footnote{The precise form of the value-stability in \cite{CIJ20} is actually $\frac{1 - x^{-2\ell}}{1 + x^{-2\ell}}$. The two functions are related to each other within small constant factors, and the form we use is mathematically more convenient for our proofs.} 

\paragraph{Simplicity through prior-free design.} 
Observe that our notion of value stability is defined over all possible pairs of value vectors $\val$ and $\val'$, and not just those that arise in a real-world context. This is intentional: as discussed in the introduction, we require the allocation algorithm to be prior-free in the sense of not knowing or even assuming a prior over value vectors. While utilizing a prior can potentially improve the algorithm's performance, there are several reasons for preferring a prior-free design: (1) prior-free algorithms are generally simpler than prior-dependent ones; (2) a single prior-free algorithm can be used in multiple different contexts without modification, and is therefore robust to changes in the market; (3) in the context of fair ad allocation, parts of the prior may be sensitive information that is inappropriate or illegal to use; and, most importantly, (4) we would like to ensure fairness guarantees even if the prior is misspecified, or if new value vectors arise that were not expected to be in the support.

\paragraph{Social welfare.}
In this work we focus on the objective of social welfare maximization. The social welfare achieved by the allocation vector $\alloc$ on value vector $\val$ is given by $\alloc \cdot \val$.  The maximum social welfare achievable for any value vector $\val$ is therefore $\max_i \vali$. As discussed previously, welfare maximizing allocation functions generally do not satisfy value stability, so we look towards approximation. In keeping with the prior-free design of the allocation function, we measure its performance in the worst case.

\begin{definition}
\label{def:approxratio}
An allocation function $\alloc$ achieves an $\compratio$-approximation to social welfare for $\compratio\le 1$ if for all value vectors $\val$ we have $\alloc(\val)\cdot \val\ge \compratio \cdot \max_i\vali$.
\end{definition}

\subsection{Fairness and value stability}
\label{sec:fairness-stability}
We now discuss how fairness connects to value stability. Chawla et al.  \cite{CIJ20} introduce a model for ad allocation where users are drawn from a universe $U$ endowed with a distance metric $d$ that captures similarities between different users: the shorter the distance between two users, the more similar they are to each other. The model employs the notion of individual fairness from \cite{DHPRZ2012} which requires, informally, that similar users should receive similar outcomes. When the outcomes are probability distributions, Dwork and Ilvento \cite{DI2018} propose using the $\ell_\infty$ distance between difference distributions as a measure of (dis-)similarity in outcomes.
\begin{definition}[Paraphrased from \cite{DHPRZ2012} and \cite{DI2018}]
  A function $\algor: U \rightarrow O$ assigning users to outcomes from a set $O$ is said to be \textbf{individually fair} with respect to distance metrics $d$ over $U$ and $D$ over $O$, if for all $u_1,u_2\in U$ we have $D(\algor(u_1), \algor(u_2)) \le d(u_1,u_2)$.

  An allocation function $\alloc:U\rightarrow \Delta([k])$ satisfies \textbf{multiple-task fairness} with respect to distance metric $d$ if for all $u_1,u_2\in U$ and $i\in [k]$, we have $|\alloci(u_1) - \alloci(u_2)| \le d(u_1,u_2)$.
\end{definition}

Like in this work, Chawla et al. \cite{CIJ20} study the design of allocation algorithms that return fair outcomes when the advertisers bid fairly. Chawla et al. \cite{CIJ20} connect fairness in advertisers' values and allocations to the distance metric over users through a constraint on advertisers' values called the ``bid ratio constraint''. Informally, if two users are close to each other under the distance metric $d$, then {\em all} advertisers are required to bid multiplicatively similar amounts on the users. \begin{definition}[\cite{CIJ20}]
A \textbf{bid ratio constraint} is a continuous function $h:[0,1] \rightarrow [1, \infty]$ with $h(0)=1$ and $h(1) = \infty$. We say that the value function $v^i: U \rightarrow [0, \infty)$ of advertiser $i$ satisfies the bid ratio constraint $h$ with respect to metric $d$ if for all $u_1,u_2\in U$: $\frac{1}{h(d(u_1,u_2))} \le \frac{v^i(u_1)}{v^i(u_2)} \le h(d(u_1,u_2))$.
\end{definition}

Chawla et al. \cite{CIJ20} then proceed to design allocation algorithms that achieve multiple task fairness with respect to the underlying metric $d$ as long as the values reported to the algorithm satisfy an appropriate bid ratio constraint. There is a close connection between our approach and that of Chawla et al. \cite{CIJ20}. An allocation algorithm is value stable if and only if it satisfies multiple task fairness whenever paired with value vectors satisfying an appropriate bid ratio constraint.
\begin{fact}
  Let $h:[0,1] \rightarrow [1, \infty]$ be a strictly increasing function with $h(0)=1$ and $h(1) = \infty$. Let $f:[1, \infty] \rightarrow [0,1]$ be the inverse of $h$. Then an allocation function $\alloc$ is value stable with respect to function $f$ if and only if over any universe $U$ endowed with distance metric $d$ and any set of value vectors assigned to users in the universe that satisfy the bid ratio constraint $h$, the function $\alloc$ applied to the value vectors produces a multiple task fair allocation with respect to $d$.
\end{fact}

\section{The Inverse Proportional Allocation algorithm and our results}\label{sec:inverseprop}

The main contribution of this paper is a new class of allocation functions that we call \InvProp (IPA) allocation. In this section we introduce the algorithm and outline its properties. Let us begin with an informal description of our mechanism. Recall that in a proportional allocation algorithm, the allocations $\alloci$ are proportional to (some function of) the values $\vali$. In the inverse proportional allocation algorithm, we set the {\em unallocated amounts} $1-\alloci$ to be proportional to {\em some decreasing function} of the values $\vali$. Specifically, the algorithm is parameterized by a decreasing function $g: \mathbb{R}^{\ge 0} \rightarrow (0, \infty]$, where $g(0) = \infty$ and $\lim_{x \rightarrow \infty} g(x) = 0$. We then deduct allocation from advertiser $i$ in proportion to $g(\vali)$.

We will now describe the algorithm more precisely in three different but equivalent ways, which will be useful for our analysis.

\vspace{0.1in}

\textbf{Formulation 1:} We start with the most direct formulation. Let $\val$ be a value vector. For $t\ge 0$ and $i\in [k]$, let $y_i(t) = \max(0, 1 - t g(\vali))$, and $y(t)=\sum_{i\in [k]} y_i(t)$. Observe that $y(t)$ is a decreasing function with $y(0)=k$. Then, there is a unique value of $t$ for which $y(t)=1$. Let $t^*>0$ be this value. Define $\alloc(\val) = (y_1(t^*), \cdots, y_k(t^*))$. (Intuitively, we start with an allocation of $1$ to every advertiser, corresponding to $t=0$. We gradually raise $t$, decreasing the allocations in proportion to $g(\vali)$. When an allocation reaches $0$, it remains $0$. The process stops when the total allocated mass is precisely equal to $1$.)

\vspace{0.1in}
\textbf{Formulation 2:} We next describe an algorithmic version that describes how to find $t^*$. For $i\in [k]$ let $w_i$ denote the unallocated amounts to be assigned to each advertiser. Observe that over any set $S$ of advertisers, the total allocated amount is at most $1$ and therefore the total unallocated amount is at least $|S|-1$. The $w_i$'s split this amount in proportion to the $g(\vali)$'s.
However, there is a problem: when the set $S$ is large and $v_i$ is very small, it may turn out to be the case that $w_i>1$, and $1-w_i$ is negative. In this case, we must assign an allocation of $0$ to the corresponding advertiser. Formally, we start with $S$ being the set of all advertisers. At each step we determine the weights $w_i(S)$ as below; we remove from $S$ any advertisers $j$ with $w_j(S)>1$; we recurse on the remaining set of advertisers until all $w_i(S)$ values are $\le 1$.
\[w_i(S) := (|S| - 1) \frac{g(v_i)}{\sum_{j \in S} g(v_j)}\]

\vspace{0.1in}
\textbf{Formulation 3:} Our final formulation is a computationally efficient algorithm implementing the above idea. See Algorithm \ref{alg:invprop}. The algorithm runs in time $O(k \log k)$. 

\begin{algorithm}[]
	\caption{Inverse Proportional Allocation (IPA) parameterized by function $g$}
	\label{alg:invprop}
	\begin{algorithmic}
		\STATE{\bfseries Input:} Function $g: \mathbb{R}^{\ge 0} \rightarrow (0, \infty]$ with $g(0) = \infty$ and $\lim_{x \rightarrow \infty} g(x) = 0$. Values $v_1, \ldots, v_k$.
		
		\STATE Initialize $\alloci = 0$ for $1 \le i \le k$

        Sort the values so that $v_1 \le v_2 \le \ldots \le v_k$.
                
        \IF {$v_k = 0$}
        \STATE Set $\alloci = 1/k$ for all $1 \le i \le k$.
        \RETURN $\alloc$.
        \ENDIF
                
        Initialize $s= \min\left(\left\{ i \in [k] \mid v_i > 0 \right\}\right)$.
                
        \WHILE {$(k-s)g(v_{s}) \ge \sum_{j=s}^k g(v_j)$ \label{step:while}}
        \STATE Increment $s$. 
        \ENDWHILE
        
        
        \FOR{$i\ge s$}
        \STATE Set $\alloci = 1 - (k - s) \frac{g(v_i)}{\sum_{j=s}^k g(v_j)}$.
        \ENDFOR
        
        \RETURN $\alloc$.
    \end{algorithmic}
\end{algorithm}

\begin{remark}
\label{remark:allocationstructure}
In Algorithm \ref{alg:invprop}, when $k = 2$, the condition for the while loop is never satisfied, so both advertisers will receive nonzero allocations if both of their values are nonzero. Similarly, whenever $k \ge 2$, at least two advertisers will be assigned nonzero allocation probabilities when there are at least two advertisers with nonzero values.
\end{remark}

\subsection{Incentive Compatibility}
Observe that IPA is a prior free algorithm. We will now formally prove incentive compatibility. The following lemma shows that the allocation assigned to any advertiser is weakly monotone increasing in her own value and weakly monotone decreasing in other advertisers' values. 
\begin{lemma}
  \label{lemma:monotonic}
For any value vector $\val$ and any $i\in [k]$, let $\val'=(\vali', \vali[-i])$ be another value vector that differs from $\val$ only in coordinate $i$ with $\vali>\vali'$. Then it holds that $\alloci(\val)\ge\alloci(\val')$ and $\alloci[j](\val)\le\alloci[j](\val')$ for all $j \neq i$.
\end{lemma}
\begin{proof}
  Consider determining the allocations $\alloc(\val)$ and $\alloc(\val')$ using Formulation 1 of the IPA described above. Let $y(t)$ and $y'(t)$ denote the sums of allocations at a particular value of $t$ for $\val$ and $\val'$ respectively. Recall that $g$ is a decreasing function, and so, $g(\vali)\le g(\vali')$. Then, at any $t$, we have $y(t)\ge y'(t)$. Therefore, the value of $t^*$ that defines the final allocations is smaller under $\val$ than under $\val'$. This immediately implies $\alloci[j](\val)\le\alloci[j](\val')$ for all $j \neq i$. Then $\alloci(\val)\ge\alloci(\val')$ follows by recalling that the allocations sum up to $1$.
\end{proof}

Lemma \ref{lemma:monotonic}, by guaranteeing monotonicity, implies that the algorithm can be implemented as an incentive-compatible mechanism using an appropriately designed payment rule. For $g(x)=1/x$, the payments can be specified explicitly; see Appendix~\ref{sec:payment}. 

\subsection{Choosing the function $g$}
An important choice in the design space for IPA algorithms is the choice of the function $g$. This choice depends on the tradeoff between the stability condition we want to satisfy and the social welfare approximation we desire. However, there is an additional property that should be satisfied. Since we focus on prior free design and worst case analysis, the algorithm should be scale free in that when all values are scaled by a common factor, the allocation remains the same.

Consider, for example, the value vectors $\val=(1,x)$ and $\valprime=(y,xy)$. The allocation produced by IPA with parameter $g$ in the two cases is, respectively,
\[ \alloc=\left( 1- \frac{g(1)}{g(1)+g(x)}, 1-\frac{g(x)}{g(1)+g(x)}\right) \,\, \text{ and } \,\,  \allocprime=\left( 1- \frac{g(y)}{g(y)+g(xy)}, 1-\frac{g(xy)}{g(y)+g(xy)}\right)  \]
In order for these allocations to be the same, modulo normalization, it must hold that $g(xy)=g(x)g(y)$ for all $x, y>0$. Under very mild assumptions on $g$ (e.g. continuity), the only solution to this functional equation is $g(x)=x^{-\ell}$ where $\ell\in (0,\infty)$. Henceforth we focus on IPA parameterized by this family of polynomial functions.

\begin{definition}
\label{def:ipa-l}
For $\ell\in (0,\infty)$, the \textbf{Inverse Proportional Allocation Algorithm with parameter $\ell$} is Algorithm~\ref{alg:invprop} with $g$ defined as $g(x)=x^{-\ell}$ for $x\in [0,\infty)$.
\end{definition}

This family of polynomial functions encompasses the entire range of tradeoffs between value stability and social welfare. To see this, consider IPA at the extremes. As $\ell \rightarrow 0$, IPA with $g(x) = x^{-\ell}$ ignores the value vector entirely and always assigns an equal allocation of $1/k$ to each advertiser. This algorithm achieves the strongest possible value stability condition, with $f(\stabpar) = 0$ for all $\stabpar \in [1, \infty]$. However, this value stability guarantees comes at the expense of social welfare: the approximation ratio approaches $0$ as $k \rightarrow \infty$. As $\ell \rightarrow \infty$, IPA with $g(x) = x^{-\ell}$ becomes a highest-bid-wins allocation rule. This algorithm achieves an approximation ratio of $1$ at the expense of value stability: $f(\stabpar)$ approaches $1$ for every $\stabpar > 1$. When $\ell \in (0, \infty)$, IPA with $g(x) = x^{-\ell}$ interpolates between these two extremes, achieving different tradeoffs between social welfare and value-stability. 

\subsection{The value stability of IPA}
We begin our analysis of the IPA by studying its value stability properties. The following theorem shows that the IPA with an appropriate choice of parameter $\ell>0$ achieves value stability with respect to the class of functions $f_\ell(\stabpar)=1-\stabpar^{-2\ell}$. 

\begin{theorem}
\label{thm:fairness}
For any $\ell>0$ and any number $k>0$ of advertisers, the inverse proportional allocation algorithm with parameter $\ell$ is value stable with respect to any function $f$ that satisfies $f(\stabpar)\ge f_\ell(\stabpar) = 1 - \stabpar^{-2\ell}$ for all $\stabpar\in [1,\infty)$.
\end{theorem}

As briefly discussed above, the family of functions $f_\ell$ essentially represents the large spectrum of possible value stability conditions. See Figure \ref{fig:theory} for a depiction of $f_{\ell}$ at different values of $\ell$. In particular, $f_0$ is $0$ everywhere and represents one extreme -- a fixed allocation; $f_\infty$ is $1$ everywhere and represents the other extreme -- an unconstrained allocation. Furthermore, for any non-decreasing function $f:[1,\infty)\rightarrow [0,1]$ with $\lim_{x\rightarrow\infty}f(x)=1$, there exists an $\ell$ with $f(x)\ge f_\ell(x)$ for all $x\in[1,\infty)$. Therefore, for any such stability constraint $f$, there exists a value for $\ell$ so that IPA with parameter $\ell$ is stable with respect to $f$.\footnote{In Section \ref{subsec:extensions}, we construct a variant of inverse proportional allocation that can better handle when $\lim_{x\rightarrow\infty}f(x) \neq 1$ and/or when the convergence to this limit is slow.}

We also show a converse of Theorem~\ref{thm:fairness}. In particular, we observe that for stability constraints $f$ that do not fall into the class of functions $\{f_\ell\}$, the best way to parameterize IPA is to choose the parameter $\ell$ corresponding to the function $f_\ell$ that most closely approximates $f$ from below.
\begin{theorem}
  \label{thm:optimalityf}
For any function $f$ and parameter $\ell>0$ such that over any number of advertisers $k>0$, the IPA with parameter $\ell$ is value stable with respect to $f$, it holds that $f(x)\ge f_\ell(x)$ for all $x\ge 1$.
\end{theorem}

The proofs of Theorem~\ref{thm:fairness} and Theorem \ref{thm:optimalityf} are deferred to Section \ref{sec:invgen}.

\subsection{The social welfare of IPA}

Next we analyze the worst case performance of the IPA for social welfare.

\begin{theorem}
\label{thm:fairvalue}
For any $\ell\in (0, \infty)$, the inverse proportional allocation algorithm with parameter $\ell$ obtains a worst case approximation ratio for social welfare of at least:
\[\compratio_\ell := \min_{x \in (0,1)} (1-x^{\ell} + x^{\ell+1}) = 1 - \frac{1}{\ell+1}\left(\frac{\ell}{\ell+1}\right)^\ell\] 
\end{theorem}

The approximation ratio of IPA is easy to analyze and we present the argument here in its entirety. The proof essentially shows that value vectors of the form $[1, x, x, \ldots, x]$ for $x < 1$ are the worst case with respect to social welfare for IPA. At such a value vector, each advertiser with value $x$ gets some non-zero allocation, which limits the mass that is placed on the highest value advertiser. 

\begin{proof}[Proof of Theorem \ref{thm:fairvalue}]
  Consider any value vector $\val$ and order the values so that $\vali[1] \ge \vali[2] \ge \ldots \ge \vali[k]$. Since multiplicative scaling does not affect the allocation or the approximation ratio, we may assume without loss of generality that $\vali[1]=1$. Then, the optimal social welfare over this value vector is also $1$. We may also assume $\vali[2]>0$, since otherwise the algorithm puts the entire allocation mass on the highest value.

  Consider the set of non-zero allocations. 
  We know that this set is of the form $\left\{1, \ldots, m \right\}$ for some $1 \le m \le k$. The social welfare achieved by IPA can be written as:
  \[\sum_{i=1}^m \vali \alloci \ge \vali[1] \alloci[1] + \vali[m] (1 - \alloci[1]) = \alloci[1] + \vali[m] (1 - \alloci[1]).  \] To analyze this expression, we first recall that $\alloci[1] = 1 - (m-1)/\sum_{i=1}^m \vali^{-\ell}$. Observe that since advertiser $m$ receives a nonzero allocation, we know that
  \[ 1\ge 1-\alloci[m]=(m-1) \frac{\vali[m]^{-\ell}}{\sum_{i=1}^m \vali^{-\ell}}, \,\, \text{ so, }\,\, \frac{m-1}{\sum_{i=1}^m \vali^{-\ell}} \le \vali[m]^{\ell}.\]
  This means that $\alloci[1] \ge 1 - \vali[m]^{\ell}$, and the approximation ratio can be lower bounded by $1-\vali[m]^{\ell} + \vali[m]^{\ell+1}$. The lowest value of this expression is achieved at $\vali[m] = \frac{\ell}{\ell+1}$, implying the bound claimed in the theorem. 
\end{proof}

Observe that the approximation achieved by IPA is independent of the number $k$ of advertisers. This is one of the main features that distinguishes IPA from the family of proportional allocation (PA) algorithms. PA necessarily exhibit worse and worse performance as the number of advertisers grows because every advertiser gets some non-zero fraction of the total allocation; IPA, on the hand, achieves good performance even when $k$ is large by better handling advertisers that have small values relative to the largest value. 

\begin{figure*}[ht]
\centering

    \begin{subfigure}{0.49\textwidth}
    \centering 
    \includegraphics[scale=0.5]{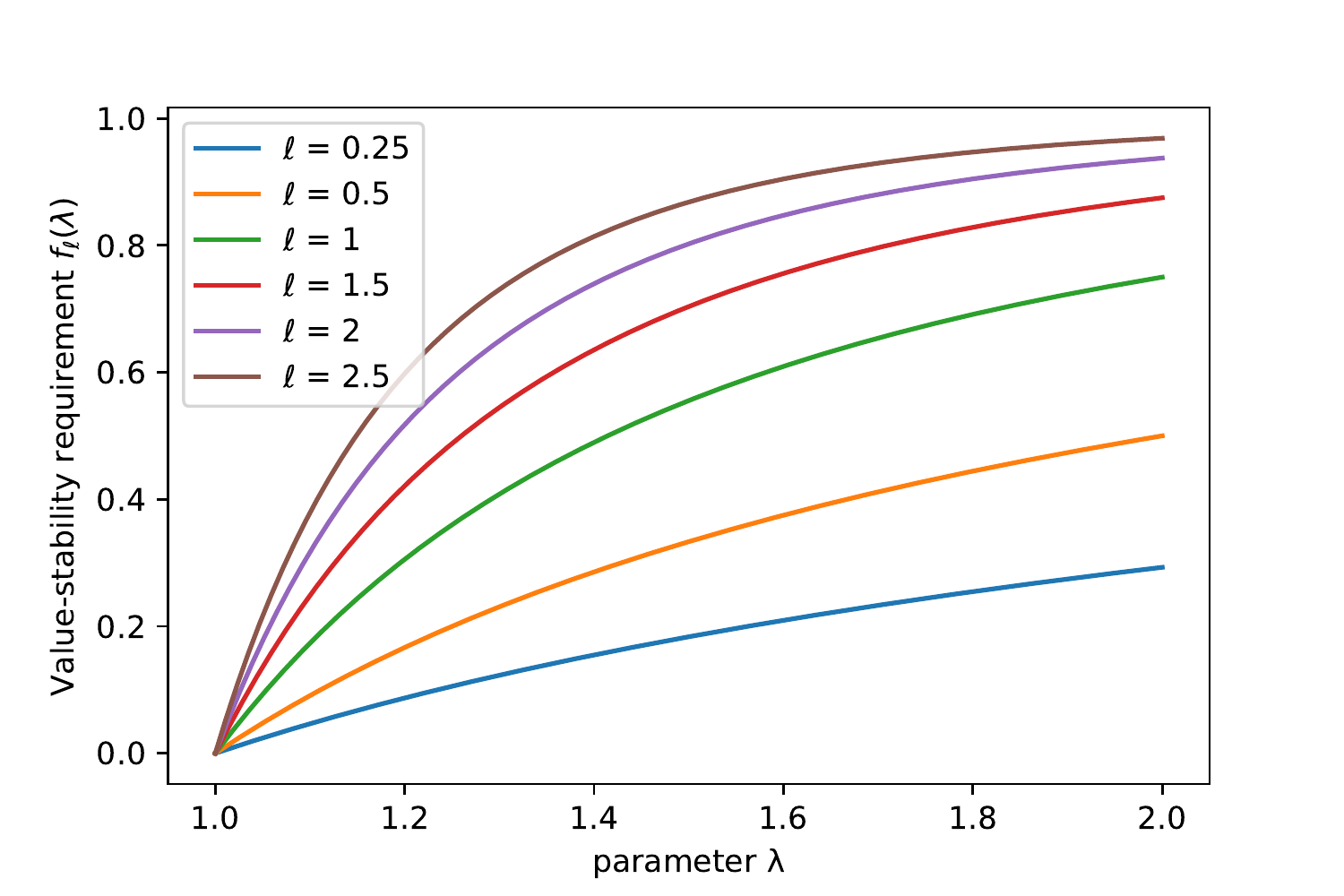}
    
    \caption{\scriptsize Theoretical analysis: value-stability conditions $f_\ell$}
    \end{subfigure}
    \begin{subfigure}{0.49\textwidth}
    \centering 
    \includegraphics[scale=0.5]{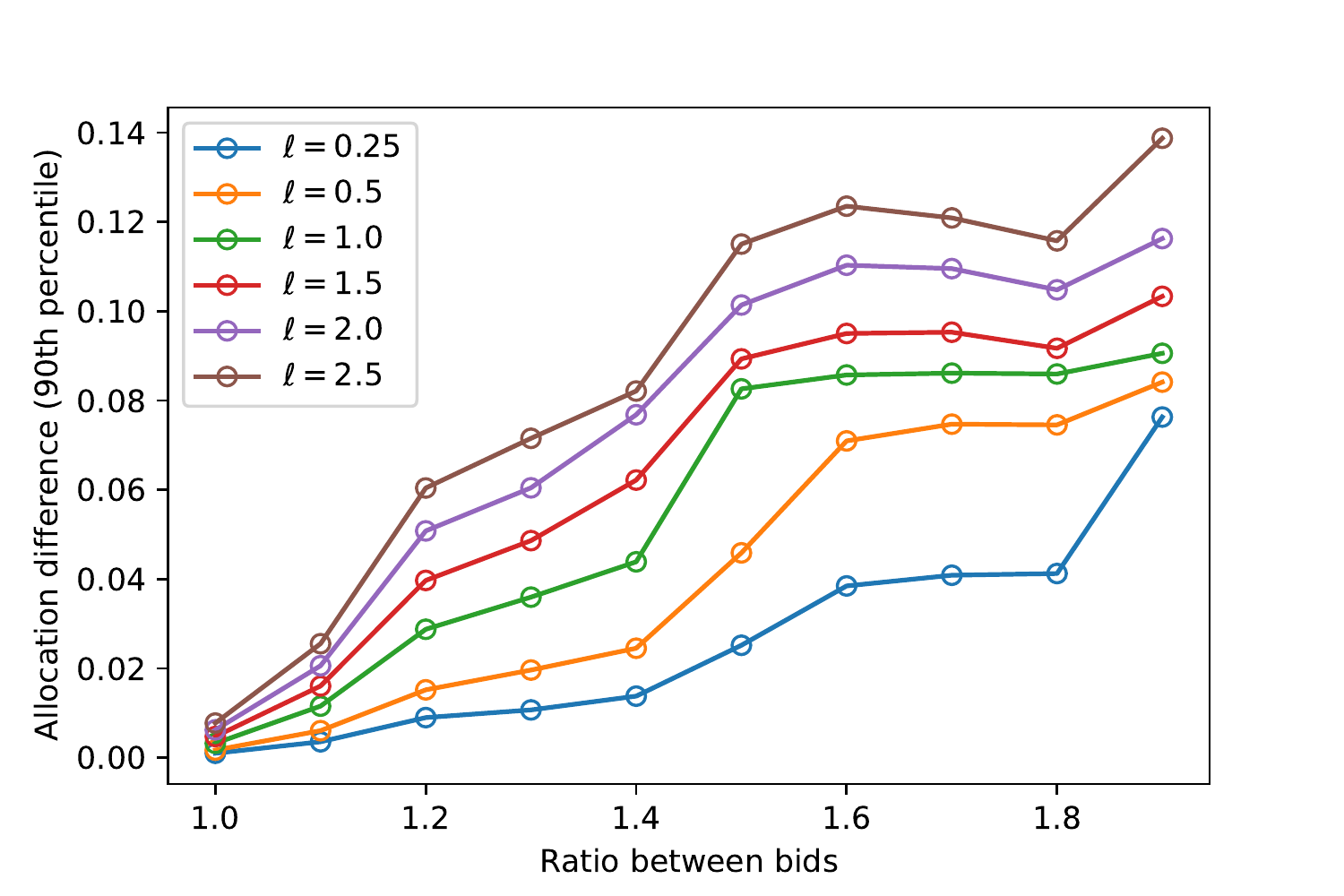}
    
    \caption{\scriptsize Empirical analysis: Bid stability profile of IPA}
    \end{subfigure}
    
        \begin{subfigure}{0.49\textwidth}
    \centering \includegraphics[scale=0.5]{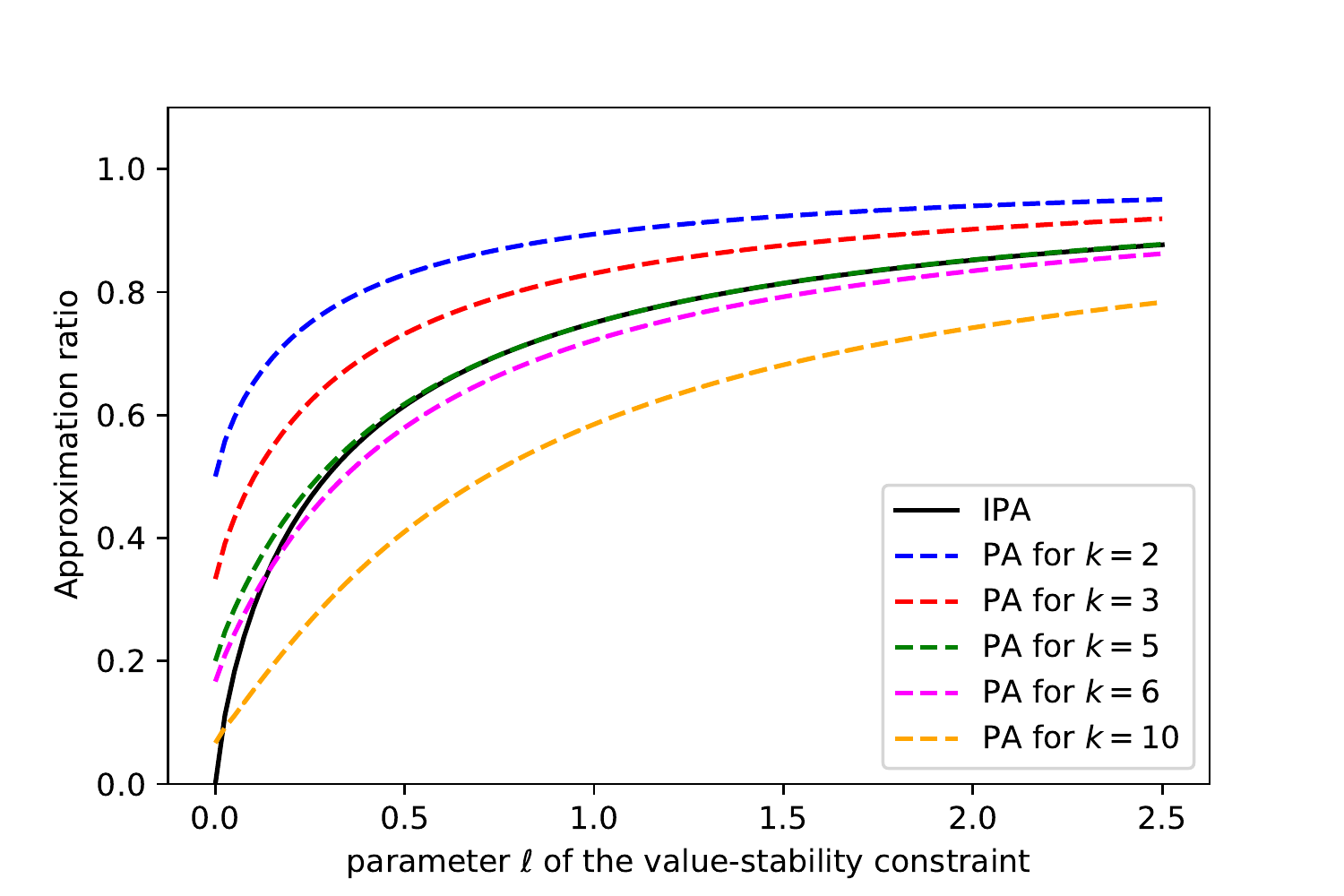}
    \caption{\scriptsize Theoretical analysis: approx ratio of IPA($\ell$), PA($2\ell$)}
    \end{subfigure}
    \begin{subfigure}{0.49\textwidth}
    \centering \includegraphics[scale=0.5]{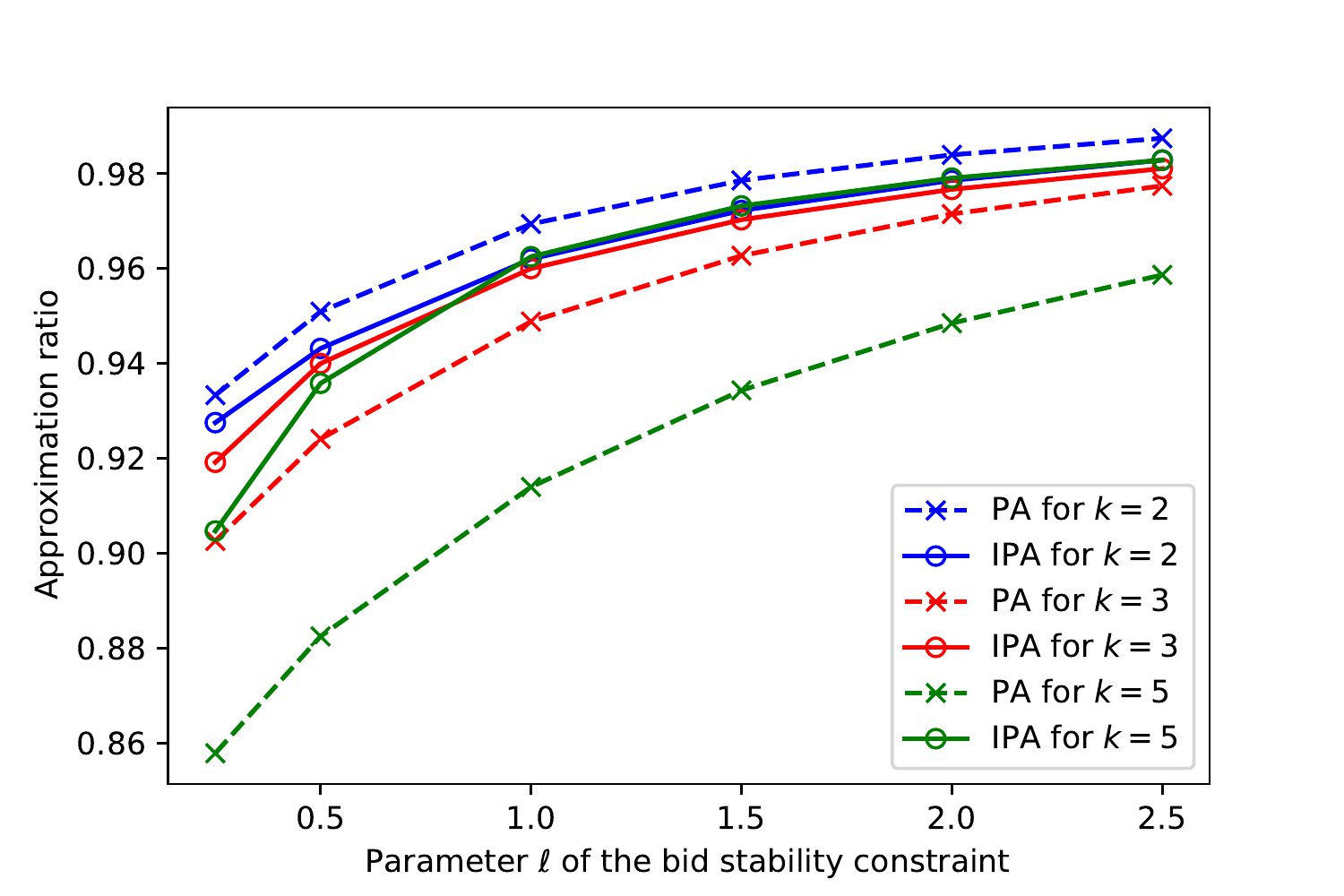}
    \caption{\scriptsize Empirical analysis: approx ratio of IPA($\ell$), PA($\frac43\ell$)}
    \end{subfigure}
    %
    \caption{Properties of IPA across different settings of $\ell$. The vertical axes are different in the first row and the second row. Refer to Section \ref{sec:experiments} for a discussion of the empirical analysis on the Yahoo dataset. (\textit{Bid-stability} in the empirical analysis is closely related to \textit{value-stability} in the theoretical analysis.)}
    \label{fig:theory}
\end{figure*}

Figure \ref{fig:theory} shows our bound from Theorem \ref{thm:fairvalue} for IPA as well as the bounds in Chawla et al. \cite{CIJ20} for PA. Notice that when $k \ge 6$, IPA consistently outperforms PA across most of the parameter regime for $\ell$. Moreover, the gap between these bounds grows as $k$ becomes larger, eventually reaching $\infty$ as $k \rightarrow \infty$. When $k = 2$, however, the bound for PA beats the bound for IPA. This turns out to be an artifact of our theoretical analysis.\footnote{For $k = 2$, the proof of Theorem \ref{thm:fairness} actually shows that the fairness guarantees of IPA turn out to be stronger, and indeed IPA with parameter $\ell$ is bid stable with respect to $f_{\ell/2}$. Thus, it is more appropriate to compare IPA with parameter $\ell$ and PA with $\ell$, and the bound in Theorem \ref{thm:fairvalue} can be improved to show that these two algorithms achieve the same approximation ratio.}


Finally, we remark that IPA with parameter $\ell=1$ arises as a full-information Nash equilibrium of an algorithm that allocates proportionally to bids with an all-pay payment rule. These mechanisms are considered in \cite{JT04, CV16, CST16}, and, in this context, the approximation ratio was previously shown to be at least $3/4$, the same bound we obtain. Our analysis is different and much simpler than those other works because, in particular, it does not require determining equilibrium strategies. Moreover, our analysis generalizes to $\ell \neq 1$.

\section{An analysis of Inverse Proportional Allocation}\label{sec:invgen}

We will now prove the value stability of Inverse Proportional Allocation, restated below.

\begin{numberedtheorem}{\ref{thm:fairness}}
For any $\ell>0$ and any number $k>0$ of advertisers, the inverse proportional allocation algorithm with parameter $\ell$ is value stable with respect to any function $f$ that satisfies $f(\stabpar)\ge f_\ell(\stabpar) = 1 - \stabpar^{-2\ell}$ for all $\stabpar\in [1,\infty)$.
\end{numberedtheorem}

\begin{numberedtheorem}{\ref{thm:optimalityf}}
  For any function $f$ and parameter $\ell>0$ such that over any number of advertisers $k>0$, the IPA with parameter $\ell$ is value stable with respect to $f$, it holds that $f(x)\ge f_\ell(x)$ for all $x\ge 1$.
\end{numberedtheorem}

To prove Theorem \ref{thm:fairness}, we need to understand how the allocation changes as the value vector changes by a small multiplicative amount. There are two main components to this analysis. First, given a value vector $\val$, we identify among all value vectors multiplicatively close to $\val$ the one that changes a particular advertiser $i$'s allocation to the greatest extent. The second step is to then bound the corresponding change to the allocation. The main challenge in doing so is that as the value vector changes, the set of advertisers that get non-zero allocation also changes. This makes it difficult to directly compare the two allocations.

Before jumping into the main proof, let us consider the special case of $k = 2$ advertisers. This setting will illustrate the first component of our analysis as well as provide intuition for how the stability constraint $f_\ell$ relates to the parameter setting of IPA. In this setting, both advertisers always receive non-zero allocation and it becomes possible to write the allocation explicitly:
\[\alloc(\val) = \left[1 - \frac{\vali[1]^{-\ell}}{\vali[1]^{-\ell} + \vali[2]^{-\ell}}, 1 - \frac{\vali[2]^{-\ell}}{\vali[1]^{-\ell} + \vali[2]^{-\ell}}\right] = \left[\frac{\vali[1]^\ell}{\vali[1]^\ell + \vali[2]^\ell}, \frac{\vali[2]^\ell}{\vali[1]^\ell + \vali[2]^\ell}\right].\]

Now fix a value $\stabpar\in [1,\infty)$ and consider any two value vectors $\val$ and $\valprime$ with $\valprimei[1] \in [1/\stabpar, \stabpar]\vali[1]$ and $\valprimei[2] \in [1/\stabpar, \stabpar]\vali[2]$. We need to show that $|\alloci[1] - \allocprimei[1]| \le f_\ell(\stabpar) = 1-\stabpar^{-2\ell}$, which implies the same bound over $|\alloci[2] - \allocprimei[2]|$. Let us fix the vector $\val$ and construct a worst-case value vector $\valprime$ such that $\allocprimei[1]$ is as small as possible while respecting the multiplicative condition on the values. The monotonicity properties given in Lemma \ref{lemma:monotonic} enable us to identify such a value vector: $\valprime= [\vali[1] / \stabpar, \stabpar \vali[2]]$. For this ``worst-case'' choice, we have:
\begin{align*}
  \alloci[1] - \allocprimei[1] &= \frac{\vali[1]^\ell}{\vali[1]^\ell + \vali[2]^\ell} - \frac{(\valprimei[1])^\ell}{(\valprimei[1])^\ell + (\valprimei[2])^\ell}\\
&= \frac{1}{1 + (\vali[2] / \vali[1])^\ell} - \frac{1}{1 + \stabpar^{2\ell} (\vali[2]/\vali[1])^\ell}.
\end{align*}
Treating $\vali[2]/\vali[1]$ as a free variable, we see that the expression is maximized at $\vali[2]/\vali[1] = 1/\stabpar$ and we get,
\[\alloci[1] - \allocprimei[1] \le \frac{1}{1 + \stabpar^{-\ell}} - \frac{1}{1 + \stabpar^\ell} = \frac{\stabpar^\ell - 1}{\stabpar^\ell + 1} \le 1 - \stabpar^{-\ell}.\]

\noindent
Recall that $f_\ell(\stabpar)=1-\stabpar^{-2\ell}$ and so the theorem follows for the case of $k = 2$.

For the general case, we carry out a similar analysis. We fix the value vector $\val$ and a parameter $\stabpar \in [1, \infty)$, and find a worst-case value vector $\valprime$ that is multiplicatively close to $\val$ but changes the allocation by the greatest amount. The challenge is that in order to write down an explicit expression for some advertiser's allocation we need to fix the set of advertisers that obtains non-zero allocation. This set changes as the value vector changes. 

Our analysis is based on two ideas. The first idea is to examine the formulation of IPA in Algorithm~\ref{alg:invprop} and consider how the allocations change if we terminate the while loop in Step~\ref{step:while} early. Let $S$ be the set of agents that receive non-zero allocations in the algorithm. We argue that if we terminate the while loop early, ``serving'' a larger set of agents than $S$, then agents in $S$ receive higher allocations. The second idea is when comparing allocations for the two vectors $\val$ and $\valprime$, if we pretend that the same set of advertisers are ``served'' under both the value vectors (with some of these allocations potentially being negative), then these allocations are additively close. We formalize these properties in the following lemmas.

\begin{lemma}
\label{prop:throwout}
Let $\val$ be any value vector with the advertisers reordered so that $\vali[1]\ge\vali[2]\ge\cdots\ge\vali[k]$.
Let $\Set=[m]$ be the set of advertisers with nonzero allocation returned by the Inverse Proportional Allocation algorithm parameterized by function $g$ on $\val$. Then for any $m'$ with $m\le m'\le k$ and any $i\in [m]$,
\[1 - (m-1)\frac{g(\vali)}{\sum_{j \in [m]} g(\vali[j])} \le 1 - (m'-1)\frac{g(\vali)}{\sum_{j \in [m']} g(\vali[j])}. \]
\end{lemma}

\begin{lemma}
  \label{prop:fairnessrestricted}
Let $\stabpar \in [1, \infty)$ and $g(x)=x^{-\ell}$ for all $x\ge 0$. Let $\Set\subseteq [k]$ be an arbitrary set of advertisers, and $i$ be an advertiser in $\Set$. Suppose that value vectors $\val$ and $\valprime$ satisfy $\valprimei = \vali / \stabpar^2$ and $\valprimei[j] = \vali[j]$ for all $j \in \Set$ with $j \neq i$. Then we have:
\[\max\left(0, 1 - (|\Set| - 1) \frac{g(\vali)}{\sum_{j \in \Set} g(\vali[j])}\right) - \max\left(0, 1 - (|\Set| - 1) \frac{g(\valprimei[i])}{\sum_{j \in \Set} g(\valprimei[j])}\right) \le f_\ell(\stabpar).\]
\end{lemma}

Lemmas \ref{prop:throwout} and \ref{prop:fairnessrestricted} along with Lemma \ref{lemma:monotonic} provide us with the necessary ingredients to prove Theorem \ref{thm:fairness}.
\begin{proof}[Proof of Theorem \ref{thm:fairness}]
  Let $\stabpar \in [1, \infty)$, and let $\val$ be an arbitrary value vector. We will focus on the allocation $\alloci[1]$ to advertiser 1, and consider a value vector $\valprime$ with $\valprimei\in [1/\stabpar,\stabpar]\vali$ for all $i\in [k]$ that assigns the minimum possible allocation to $1$. Since the IPA is a scale-free allocation algorithm, we may equivalently focus on value vectors $\valprime$ with $\valprimei\in [1/\stabpar^2,1]\vali$ for all $i\in [k]$. Lemma \ref{lemma:monotonic} then implies that the allocation to $1$ is minimized at $\valprimei[1]=\vali[1]/\stabpar^2$ and $\valprimei[-1]=\vali[-1]$.

  Let $\Set$ be the set of advertisers receiving nonzero allocation from IPA on $\val$, and let $\Setprime$ be the corresponding set for $\valprime$. Let $\alloc$ and $\allocprime$ be the corresponding allocations. By Lemma \ref{lemma:monotonic}, $\allocprimei\ge\alloci$ for all $i\ne 1$ and $\allocprimei[1]\le\alloci[1]$. Therefore, $\Set\subseteq \Setprime\cup\{1\}$. We want to show: $\alloci[1]-\allocprimei[1]\le f_\ell(\stabpar)$. We now break up our analysis into three cases.

\textbf{Case 1: $1 \not\in \Set$.} In this case, $\alloci[1]=0$ and $\alloci[1] - \allocprimei[1]\le 0$, so the statement trivially holds.

\textbf{Case 2: $1\in\Set$ and $1 \in \Setprime$.}
In this case we have $\Set\subseteq\Setprime$, and $\alloci[1], \allocprimei[1]>0$. So, by Lemma \ref{prop:throwout} we obtain: 
\[\alloci[1] = 1 - (|\Set| - 1)\frac{g(\vali[1])}{\sum_{j \in \Set} g(\vali[j])} \le 1 - (|\Setprime| - 1)\frac{g(\vali[1])}{\sum_{j \in \Setprime} g(\vali[j])}.\] Furthermore, Lemma \ref{prop:fairnessrestricted} implies:
\[1 - (|\Setprime| - 1)\frac{g(\vali[1])}{\sum_{j \in \Setprime} g(\vali[j])}\le \max\left(0,1 - (|\Setprime| - 1)\frac{g(\valprimei[1])}{\sum_{j \in \Setprime} g(\valprimei[j])} \right) + f_\ell(\stabpar) = \allocprimei[1]  + f_\ell(\stabpar) ,\] so $\alloci[1] - \allocprimei[1]\le f_\ell(\stabpar)$ as desired.

\textbf{Case 3: $1\in\Set$ and $1 \not\in \Setprime$.} In this case we cannot directly apply Lemma~\ref{prop:fairnessrestricted} to $\Setprime$. Instead,
let $\Rdoubleprime = \left\{i \mid \valprimei \ge \valprimei[1] \right\} = \left\{i \mid \vali \ge \vali[1] / \stabpar^2 \right\}$. We claim that $S \subseteq \Rdoubleprime$. This is because 
$1 \not\in \Setprime$, and therefore all $i\in\Setprime$ have $\valprimei\ge\valprimei[1]$, and therefore $\Set\subseteq\Setprime \cup \left\{1\right\} \subseteq \Rdoubleprime$.

Thus, we can apply Lemma~\ref{prop:throwout} to obtain: 
\[\alloci[1] = 1 - (|\Set| - 1)\frac{g(\vali[1])}{\sum_{j \in \Set} g(\vali[j])} \le 1 - (|\Rdoubleprime| - 1)\frac{g(\vali[1])}{\sum_{j \in \Rdoubleprime} g(\vali[j])}. \]

Next, we can apply Lemma \ref{prop:fairnessrestricted} to $\Rdoubleprime$ to see that 
\[1 - (|\Rdoubleprime| - 1)\frac{g(\vali[1])}{\sum_{j \in \Rdoubleprime} g(\vali[j])} \le \max\left(0,1 - (|\Rdoubleprime| - 1)\frac{g(\valprimei[1])}{\sum_{j \in \Rdoubleprime} g(\valprimei[j])}\right) + f_\ell(\stabpar).\] 

Finally, since advertiser 1 receives a zero allocation on value vector $\valprime$ and since $\valprimei[1] = \min \left(\left\{\valprimei[i] \mid i \in \Rdoubleprime \right\} \right)$, we know by Step \ref{step:while} of Algorithm \ref{alg:invprop} that $(|\Rdoubleprime| -  1) g(\valprimei[1]) \ge \sum_{j \in \Rdoubleprime} g(\valprimei[j])$. Thus, 
$1 - (|\Rdoubleprime| -  1)\frac{g(\valprimei[1])}{\sum_{j \in \Rdoubleprime} g(\valprimei[j])} \le 0$. This shows that $\alloci[1] \le f_\ell(\stabpar) = \allocprimei[1] + f_\ell(\stabpar)$ as desired. 
\end{proof}

\noindent
It remains to prove the lemmas.
\begin{proof}[Proof of Lemma~\ref{prop:throwout}]
For $i \in [m]$, let $\alloci = 1 - (m-1)\frac{(1/\vali)}{\sum_{j \in [m]} (1/\vali[j])}$ and $\allocprimei[i] = 1 - (m'-1)\frac{(1/\vali)}{\sum_{j \in [m']} (1/\vali[j])}$. We wish to prove that for $i \in [m]$, it holds that $\alloci \le \allocprimei$. In order to compare these values, we consider the following expressions for $\alloci$ and $\allocprimei$  using Formulation 1 of IPA in Section \ref{sec:inverseprop}. First, we see that $\alloci[j] = \max(0, 1 - \algpar \cdot (1/\vali[j]))$, where $\algpar$ is the unique value such that $\sum_{j=1}^k \alloci[j] = 1$. Equivalently, since $[m']$ contains the set of advertisers that receive nonzero allocations, $t$ is also the unique value such that $\sum_{j \in [m']} \alloci[j] = 1$. Similarly, we see that $\allocprimei[j] = 1 - \algprimepar \cdot (1/\vali[j])$, where $\algprimepar$ is the unique value such that $\sum_{j \in [m']} \alloci[j] = 1$. In other words, the only difference between $\algpar$ and $\algprimepar$ is that the allocations are not constrained to be nonnegative for $\algprimepar$. This implies that $\algprimepar \le \algpar$, and so for $i \in [m]$:
\[\alloci = \max(0, 1 - \algpar \cdot(1/\vali)) = 1 - \algpar \cdot (1/\vali) \le 1 - \algprimepar \cdot (1/\vali) = \allocprimei[i]\] as desired. 
\end{proof}

\begin{proof}[Proof of Lemma~\ref{prop:fairnessrestricted}]
Let us consider the expression we need to bound:

\begin{align*}
 d & := \max\left(0, 1 - (|\Set| - 1) \frac{g(\vali)}{\sum_{j \in \Set} g(\vali[j])}\right) - \max\left(0, 1 - (|\Set| - 1) \frac{g(\valprimei[i])}{\sum_{j \in \Set} g(\valprimei[j])}\right) \\
  & = 
      \max\left(0, 1 -  \frac{(|\Set| - 1) \cdot g(\vali)}{g(\vali)+\sum_{j \in \Set; j\ne i} g(\vali[j])}\right) - \max\left(0, 1 -  \frac{(|\Set| - 1) \cdot g\left(\frac{\vali}{\stabpar^2}\right)}{g\left(\frac{\vali}{\stabpar^2}\right)+\sum_{j \in \Set; j\ne i} g(\vali[j])}\right)
\end{align*}

We want to show that $d\le f_\ell(\stabpar)$. We first observe that the statement is vacuously true when the maximum in the first term of $d$ is equal to $0$ as well as when $|S|=1$. So henceforth we assume that $|S|\ge 2$ and $1 - (|\Set| - 1) \frac{g(\vali)}{g(\vali)+\sum_{j \in \Set; j\ne i} g(\vali[j])} > 0$.

Let $\sumpar$ denote the ratio $(\sum_{j \in \Set, j \neq i} g(\vali[j]))/g(\vali)$. Recalling that $g(\vali/\stabpar^2) = \stabpar^{2\ell}g(\vali)$, we get,
\begin{align*}
  d(\sumpar) &= 1 - \frac{|\Set| - 1}{1+\sumpar} - \max\left(0, 1 - \frac{|\Set| - 1}{1+\stabpar^{-2\ell}b}\right)\\
  & = \frac{|\Set| - 1}{\max(|\Set|-1, 1+\stabpar^{-2\ell}b)} - \frac{|\Set| - 1}{1+\sumpar}
\end{align*}
Now we will consider three cases. First, if $|S|=2$, the expression becomes
\begin{align*}
  d(\sumpar) & = \frac{1}{1+\stabpar^{-2\ell}b} - \frac{1}{1+\sumpar}
\end{align*}
This is identical to the expression we analysed for the $2$ advertisers setting at the beginning of this section. The expression maximized at $b=\stabpar^\ell$, and we obtain
\begin{align*}
  d(\sumpar) &\le \frac{\stabpar^\ell - 1}{\stabpar^\ell + 1} \le 1 - \stabpar^{-\ell}.
\end{align*}
Second, suppose that $|S|>2$ and $\max(|\Set|-1, 1+\stabpar^{-2\ell}\sumpar)=|\Set|-1$. This means that $\sumpar \le \stabpar^{2\ell} (|\Set| - 2)$. Then we have, $d(\sumpar) = 1- (|\Set| - 1)/(1+\sumpar)$. Note that $d(\sumpar)$ is increasing as a function of $\sumpar$. So in this case it is maximized at $\sumpar =\stabpar^{2\ell} (|\Set| - 2)$. We get
\begin{align*}
  \max_{\sumpar \le \stabpar^{2\ell} (|\Set| - 2)} d(\sumpar) &\le 1- \frac{|\Set| - 1}{1+\stabpar^{2\ell} (|\Set| - 2)} \le 1-\stabpar^{-2\ell} = f_\ell(\stabpar)
\end{align*}
Finally, suppose that $|S|>2$ and $\max(|\Set|-1, 1+\stabpar^{-2\ell}\sumpar)=1+\stabpar^{-2\ell}\sumpar$. This means that $\sumpar \ge \stabpar^{2\ell} (|\Set| - 2)$. Then we have
\begin{align*}
  d(\sumpar) & = (|\Set|-1) \left(\frac{1}{1+\stabpar^{-2\ell}\sumpar}-\frac{1}{1+\sumpar}\right)
\end{align*}
In this case, the function $d(\sumpar)$ is non-increasing on $\sumpar\in [\stabpar^{2\ell}(|\Set| - 2), \infty)$ (see Proposition \ref{prop:derivative}.) Therefore, once again we have,
\begin{align*}
  \max_{\sumpar \ge \stabpar^{2\ell} (|\Set| - 2)} d(\sumpar) &\le 1- \frac{|\Set| - 1}{1+\stabpar^{2\ell} (|\Set| - 2)} \le 1-\stabpar^{-2\ell} = f_\ell(\stabpar)
\end{align*}
This completes the proof.
\end{proof}

We now prove Theorem \ref{thm:optimalityf}.
\begin{proof}[Proof of Theorem \ref{thm:optimalityf}]
Let $x$ be any value in $[1, \infty]$. We consider the value vectors $\val = [x, \ldots, x]$ and $\valprime = [x^2, 1, \ldots, 1]$. Notice that $\max_{i \in [k]} \left(\max\left(\frac{\vali}{\valprimei}, \frac{\valprimei}{\vali} \right)\right) = x$. Observe that the allocation $\allocprimei[1]$ is $1 - (k-1)\frac{g(x^2)}{g(x^2) + (k-1) g(1)} = 1 - \frac{g(x^2)}{\frac{g(x^2)}{k-1} + g(1)}$. As $k \rightarrow \infty$, this becomes $1 - \frac{g(x^2)}{g(1)} \rightarrow 1 - x^{-2\ell}$. In comparison, the allocation $\alloci[1] \rightarrow 0$. Thus, we must have that $f(x) \ge 1 - x^{-2\ell}$. 
\end{proof}

\section{Near-optimality of Inverse Proportional Allocation}\label{sec:genf}

In this section, we prove that inverse proportional allocation is near-optimal in comparison to any prior-free allocation algorithm. First, we show in Section \ref{subsec:optimality} that IPA is optimal for the family of value-stability constraints $f_{\ell}$: that is, no prior-free allocation algorithm can beat the performance of IPA within this family of value-stability constraints. Then, we consider the performance of IPA for value-stability constraints $f$ that are not of the form $f_{\ell}$. We construct an extension of IPA called {\em capped IPA} that performs near-optimally for any ``reasonable'' function $f$ satisfying a mild constraint. In Section \ref{subsec:extensions}, we discuss this extension of IPA, and in Section \ref{subsec:nearoptimality}, we prove our near-optimality result: for any function $f$ satisfying the constraint that admits a prior-free value-stable allocation algorithm with approximation ratio $\compratio$,  there exists a capped IPA algorithm that achieves an approximation ratio of $\Omega(\frac{\compratio}{\ln(1/\compratio)+1})$.

\subsection{Optimality of IPA}
\label{subsec:optimality}

Recall that Theorems~\ref{thm:fairness} and \ref{thm:fairvalue} together imply that for any $\ell>0$, it is possible to achieve value stability with respect to the function $f_\ell$ while obtaining an $\compratio_\ell = 1-\ell^\ell/(\ell+1)^{\ell+1}$ approximation to social welfare. We will now show that no prior-free allocation algorithm that is value stable with respect to $f_\ell$ can obtain a better approximation as the number of advertisers $k$ tends to infinity.

\begin{theorem}
  \label{thm:optimalityfl}
  For every $\ell>0$, no prior-free allocation algorithm that is value stable with respect to $f_\ell$ can obtain a worst case approximation ratio for social welfare over $k$ advertisers that is better than $\compratio_\ell-1/k$.
\end{theorem}
\noindent Thus, within the family $f_{\ell}$, the approximation ratio obtained by IPA is optimal within an additive factor of $1 / k$. In the limit as $k \rightarrow \infty$, this theorem implies that IPA achieves the optimal approximation ratio in comparison to any prior-free allocation algorithm that is value-stable for $f_{\ell}$.

The proof of this theorem is based on the following lemma from \cite{CIJ20} that we restate and reprove for completeness.
\begin{lemma}[\cite{CIJ20}, rephrased]
\label{lemma:CIJupperbound}
Let $f: [1,\infty] \rightarrow [0,1]$ be any weakly increasing function, and $\stabpar>1$ be an arbitrary real number. Then, no prior free allocation algorithm over $k$ advertisers that is value-stable with respect to $f$ can obtain an approximation to social welfare better than:
$\left(\frac{1}{k} + f(\stabpar) + \stabpar^{-2} \left(1 - \frac{1}{k} -f(\stabpar \right)\right)$. 
\end{lemma}
\begin{proof}
 Consider the value vectors $\val = [1, \ldots, 1]$. Since the total allocation is $1$, there must exist an advertiser $i \in [k]$ such that $\alloci[i] \le 1/k$. We define a value vector $\valprime$ that's within a multiplicative factor of $\stabpar$ on every coordinate and place a lower bound on its approximation ratio for social welfare. More specifically, we consider $\valprime$ so $\valprimei[i] = \stabpar$ and $\valprimei[j] = \stabpar^{-1}$ for $j \neq i$. Since the allocation is value-stable, it must satisfy $\allocprimei[i] \le f(\stabpar) + \alloci[i] \le \frac{1}{k} + f(\stabpar)$. Thus, the approximation ratio for social welfare is $\allocprimei[i] + \stabpar^{-2} (1 - \allocprimei[i])$. Using the above bound on $\allocprimei[i]$, we see that the approximation ratio is at most $\frac{1}{k} + f(\stabpar) + \stabpar^{-2} \left(1 - \frac{1}{k} -f(\stabpar) \right)$ as desired. 
\end{proof}

The proof of Theorem~\ref{thm:optimalityfl} now follows directly.  

\begin{proof}[Proof of Theorem \ref{thm:optimalityfl}]
 The theorem follows from Lemma~\ref {lemma:CIJupperbound} by substituting $f(\stabpar)=1-\stabpar^{-2\ell}$; minimizing the expression over $\stabpar$; and recalling that $\compratio_\ell = \min_{x\in (0,1)} (1-x^\ell+x^{\ell+1})$, which can be written as $\min_{\stabpar> 1}\left (1-\stabpar^{-2\ell}+\stabpar^{-2(\ell+1)}\right) = \min_{\stabpar> 1}\left( f(\stabpar) +\stabpar^{-2}(1 - f(\stabpar) \right)$. 
\end{proof}

\subsection{A simple extension of IPA}\label{subsec:extensions}
We now consider general value-stability constraints $f$, and consider the design of algorithms that are value-stable for general $f$. As mentioned earlier, for every function $f$ with $\lim_{x\rightarrow\infty} f(x)=1$, there exists a parameter $\ell>0$ such that $f(x)\le f_\ell(x)$ for all $x\in [1,\infty)$, and so value stability for such functions can be achieved with an IPA algorithm. However, this family excludes stability constraints $f$ with $\lim_{x\rightarrow\infty} f(x)=\multpar<1$. With this shortcoming in mind, we propose an extension of IPA that we call {\em capped IPA}. The idea is to  run IPA with probability $\multpar$, so that the maximum difference between any two advertiser's allocations is always capped by $\multpar$.

\begin{alg}
\label{alg:cappedIPA}
The \textbf{Capped Inverse Proportional Allocation algorithm with parameters $\ell$ and $\multpar$} applies the Inverse Proportional Algorithm with parameter $\ell$ with probability $\multpar$ and uniformly assigns allocation across all advertisers with probability $1 - \multpar$.
\end{alg}

The value stability and social welfare properties of capped inverse proportional allocation follow immediately from our analysis of IPA in the previous sections. 

\begin{corollary}
\label{cor:cappedIPA} 
Let $\ell>0$ and $\multpar \in [0, 1]$ be parameters, and let $k>0$ be any number advertisers. The capped inverse proportional allocation algorithm with parameters $\ell$ and $\multpar$ is value stable with respect to any function $f$ that satisfies $f(\stabpar)\ge \multpar f_\ell(\stabpar) = \multpar (1 - \stabpar^{-2\ell})$ for all $\stabpar\in [1,\infty)$. Moreover, this algorithm obtains a worst case approximation ratio for social welfare of at least
\[\multpar \cdot \compratio_\ell = \multpar \left(1 - \frac{1}{\ell+1}\left(\frac{\ell}{\ell+1}\right)^\ell\right).\]
\end{corollary}
\begin{proof}
For any value vector $\val$, we compare the allocation $\alloc$ of this capped IPA algorithm to the allocation $\allocprime$ of IPA with parameter $\ell$. For any $i \in [k]$, it holds that $\alloci[i] = \multpar \allocprimei[i] + (1 - \multpar)/k$. The value-stability and social welfare properties then follow from Theorem \ref{thm:fairness} and Theorem \ref{thm:fairvalue}. 
\end{proof}

\subsection{Near-optimality with respect to general value-stability functions}\label{subsec:nearoptimality}

We now show that capped inverse proportional allocation can perform near-optimally for any ``reasonable'' value stability constraint $f:[1, \infty] \rightarrow [0, 1]$. First, we argue that any ``reasonable'' value-stability constraint $f$ should at minimum satisfy the following conditions: (a) $f$ is weakly increasing, and (b) $f(x) / \ln(x)$ is weakly decreasing.

For constraint (a), observe that if a prior-free allocation algorithm is value-stable with respect to $f$, then it is also value-stable with respect to $g(x) = \min_{y \ge x} g(y)$ which is weakly increasing.

For constraint (b), observe that if a prior-free allocation algorithm is value-stable with respect to $f$, then it is also value-stable with respect to $g(x) = \inf_{n \in \mathbb{N}} nf(x^{1/n})$. Then for any $x\in [1,\infty]$ and $n \in \mathbb{N}$, it holds that
$g(x) \cdot n = n \cdot \inf_{m \in \mathbb{N}} mf(x^{1/m}) \ge \inf_{m' \in \mathbb{N}} m' f(x^{n/m'})  = g(x^n)$. This implies that $\frac{g(x)}{\ln (x)}  \ge \frac{g(x^n)}{\ln(x^n)} $. 
After some ``smoothing'', this essentially guarantees that $g(x) / \ln(x)$ is weakly decreasing.  

With these conditions in place, we consider value-stability constraints $f$ where $f$ is weakly increasing and $f(x) / \ln(x)$ is weakly decreasing. We now show that for any $f$ of this form, the family of capped inverse proportional allocation algorithms achieves value-stability while obtaining an approximation ratio that is only logarithmically worse than optimal. 

\begin{theorem}
\label{thm:nearoptimality}
Let $f: [1, \infty] \rightarrow [0,1]$ be any function where $f(x) / \ln(x)$ is weakly decreasing and $f$ is weakly increasing. Suppose that there exists a prior-free allocation algorithm that is value-stable for $f$ that achieves a worst-case approximation ratio of $\compratio$. Then, there exist parameters $\ell \in (0, \infty)$ and $\multpar \in [0,1]$ such that the capped IPA algorithm with parameters $\ell$ and $\multpar$ is value-stable for $f$ and achieves a worst-case approximation ratio of at least: $\left(\frac{\compratio/2}{2\ln\left(1/\compratio\right)+1}\right)$. 
\end{theorem}

The proof of Theorem \ref{thm:nearoptimality} leverages Lemma~\ref{lemma:CIJupperbound} to lower bound $f(x)$ as a function of the approximation ratio. This bound is applied to a carefully chosen point, and the basic properties of $f$ can be used to extend this to a pointwise lower bound by a ``well-behaved'' function $F$. This function $F$ can be approximated from below by a function of the form $\multpar \cdot f_{\ell}$. We then show that capped IPA with parameters $\ell$ and $\multpar$ achieves a significant fraction of the optimal approximation ratio for the value-stability constaint $f$.

\begin{proof}[Proof of Theorem \ref{thm:nearoptimality}]
Let $\multpar = \frac{\compratio-\compratio^2}{1-\compratio^2} = \frac{\compratio}{1+\compratio}$, and let $F(x) = \min(\multpar, \ln(x) \cdot \multpar / \ln(1/\compratio))$. We begin by showing that $f(x) \ge F(x)$ for all $x \in [1, \infty]$. By Lemma \ref{lemma:CIJupperbound}, we know that $f(x) \ge (\compratio-x^{-2})/(1-x^{-2})$ for every $x \in [1, \infty)$. Now, we apply this fact at $x = 1 / \compratio$ to conclude that $f(1 / \compratio) \ge \frac{\compratio-\compratio^2}{1-\compratio^2} = \multpar$. Using the basic properties of $f$, we can extend this to obtain a pointwise lower bound by $F$. Since $\frac{f(x)}{\ln(x)}$ is decreasing, this means that for $x \le 1/\compratio$, we have that $f(x) \ge \ln(x) \cdot \multpar / \ln(1/\compratio) = F(x)$. Moreover, since $f$ is weakly increasing, we know that for $x \ge 1/\compratio$, we have that $f(x) \ge \multpar = F(x)$. Putting these two conditions together, we have that $f(x) \ge \min(\multpar, \ln(x) \cdot \multpar / \ln(1/\compratio)) = F(x)$. 

Now, we relate $F(x)$ to the value-stability condition for a capped IPA algorithm. By Proposition \ref{prop:boundinglogs}, we know that $\min(\multpar, \ln(x) \cdot \multpar / \ln(1/\compratio)) \ge \multpar (1 - x^{-1/\ln(1/\compratio)})$. This implies that $f(x) \ge \multpar (1 - x^{-1/\ln(1/\compratio)})$ as desired. Let $\ell = 1/(2\ln(1/\compratio))$. We claim that the capped IPA with parameters $\ell$ and $\multpar$ is value-stable for $f$. By Corollary \ref{cor:cappedIPA}, we know that this capped IPA algorithm is value-stable for $f(x) \ge \multpar(1 - x^{-2\ell}) = \frac{\compratio-\compratio^2}{1-\compratio^2} (1 - x^{-1/\ln(1/\compratio)})$, as desired. 

Now, we show that this capped IPA algorithm achieves an approximation ratio of at least $\left(\frac{\compratio/2}{2\ln\left(1/\compratio\right)+1}\right)$. By Corollary \ref{cor:cappedIPA}, we know that the algorithm achieves social welfare of at least 
\[\multpar\left(1 - \frac{1}{\ell+1} \left(\frac{\ell}{\ell+1}\right)^{\ell}\right) \ge \frac{\compratio - \compratio^2}{1-\compratio^2} \left(1 - \frac{1}{\ell+1}\right) = \frac{\compratio}{1 + \compratio} \left(\frac{1}{2 \ln(1/\compratio) + 1} \right). \] This completes the proof.
\end{proof}

Theorem \ref{thm:nearoptimality} shows that capped IPA achieves an approximation ratio of at least  $\frac{\compratio/2}{2\ln\left(1/\compratio\right)+1}$, where $\compratio$ is the best possible approximation ratio achieved by any prior-free allocation algorithm that is value-stable for $f$. In fact, this result also shows that when $\compratio \ge 0.5$ (i.e. the best possible approximation ratio achieves a $0.5$ fraction of the social welfare), then capped IPA achieves a approximation ratio of at least $0.2 \compratio$. In other words, in this regime, capped IPA is able to achieve a \textit{constant fraction} of the optimal approximation ratio. When the approximation ratio is small, the guarantees provided in Theorem \ref{thm:nearoptimality} do degrade with $\ln(1/\compratio)$. An interesting direction for future work would be to try to close this gap and maintain a constant fraction of the optimal social welfare even when the approximation ratio is small.

\section{Experiments}\label{sec:experiments}

We empirically evaluated the performance of our proposed ad auction algorithm, IPA, as well as that of PA and the highest-bid-wins auction on the Yahoo A1 dataset \cite{Yahoo}. Our experiments show that the performance of these algorithms aligns with our theoretical findings: (1) the highest-bid-wins auction does not achieve fairness on this dataset; (2) IPA achieves strong fairness and social welfare guarantees; and (3) for an appropriate comparison of parameters, IPA can achieve better tradeoffs between fairness and social welfare than PA, particularly when the number of advertisers is large. 

\paragraph{Experimental Setup.} The Yahoo Search Marketing Advertiser Bidding Dataset (A1) \cite{Yahoo} contains sponsored search auction data from Yahoo's platform for the top 1000 search queries between 2002 and 2003. Advertisers compete to appear alongside search results for each keyword, and the dataset contains bidding information in 15 minute time increments. During the years in which the data was collected, Yahoo utilized a first-price auction format. This dataset has been used in previous work on fairness in ad auctions \cite{celis2019, NTC2020}. 

We pre-processed the Yahoo A1 dataset to retain all keywords with at least $2$ bids. Since the dataset does not contain fine-grained time information, we treated all bids on the same keyword within a 15 minute time increment as participating in the same ad auction;\footnote{In the case of multiple bids from a single advertiser, we consider only their most recent bid.} the corresponding keyword is treated as a unique user. We then considered all pairs of keywords within a certain \textit{time-horizon}, measured their similarity, and evaluated the fairness of different allocation algorithms over these pairs. The choice of the time-horizon can be consequential. In order to study performance over long enough intervals as well as ensure robustness of our results over the entire duration, we partitioned the dataset into 11 time horizons, each of length one month.

In these experiments, we do not distinguish between bids and values. Since the dataset provides advertiser bids, we refer to the fairness requirement as \textit{bid-stability}, rather than value-stability. The approximation ratio is similarly computed using the bids rather than the values.


\begin{figure}[h]
    \centering
    \includegraphics[scale=0.6]{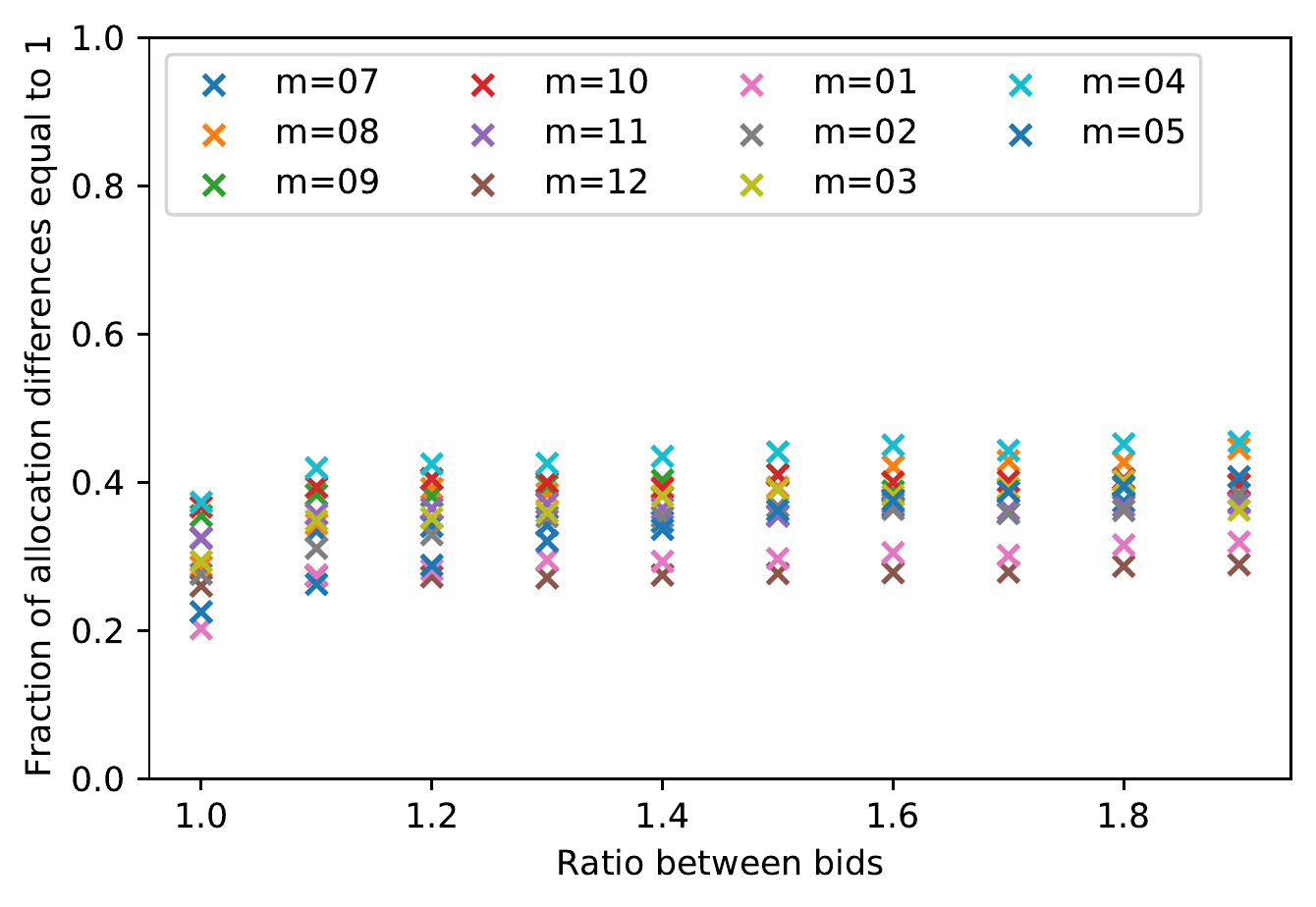}
    \caption{Bid-stability analysis of the highest-bid-wins auction}
    \label{fig:highestbid}
\end{figure}

\begin{figure*}[ht]
    \centering
    \begin{subfigure}{0.49\textwidth}
    \includegraphics[scale=0.52]{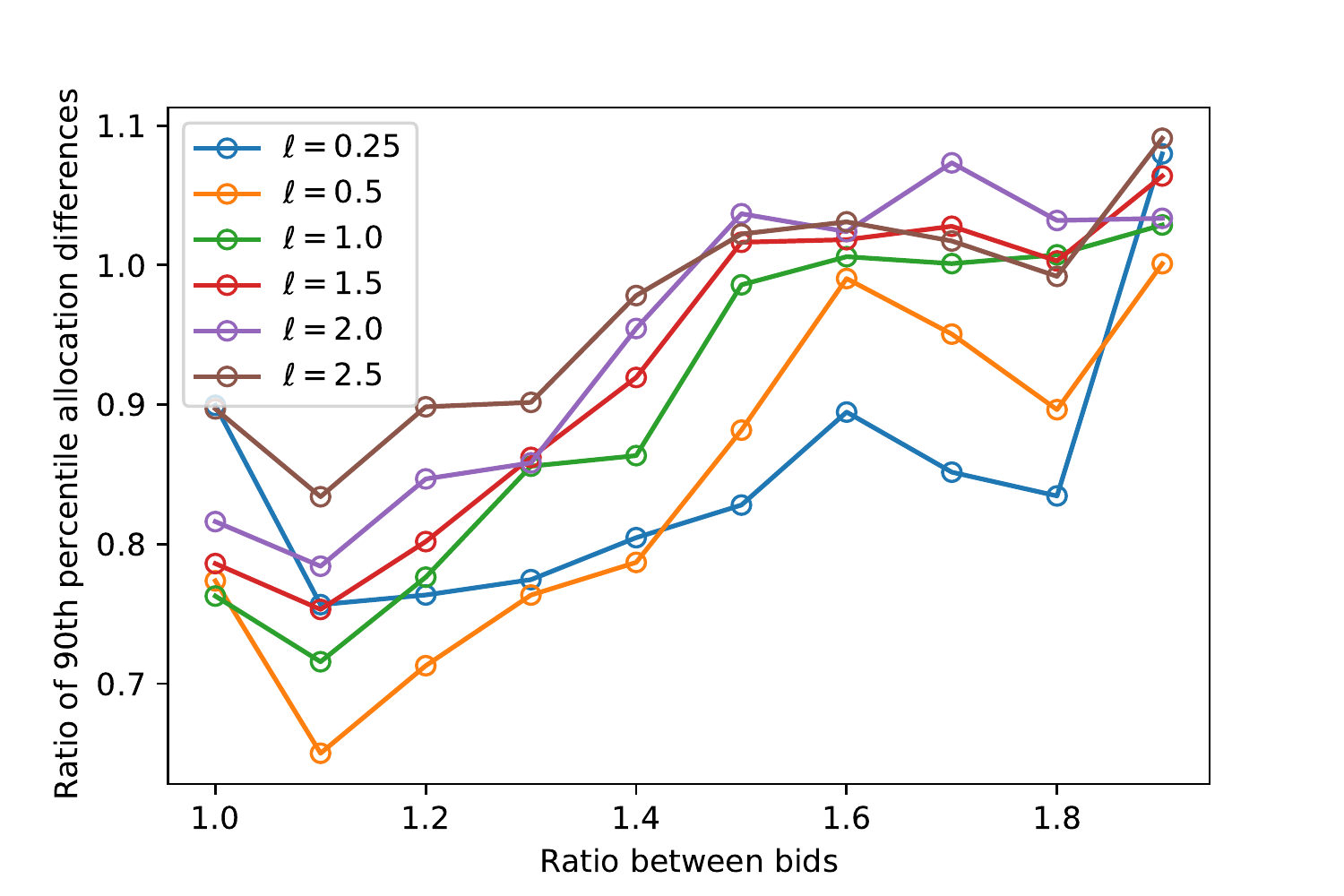}
    \caption{Ratios of bid stability profiles} 
    \end{subfigure}
    \begin{subfigure}{0.49\textwidth}
    \includegraphics[scale=0.52]{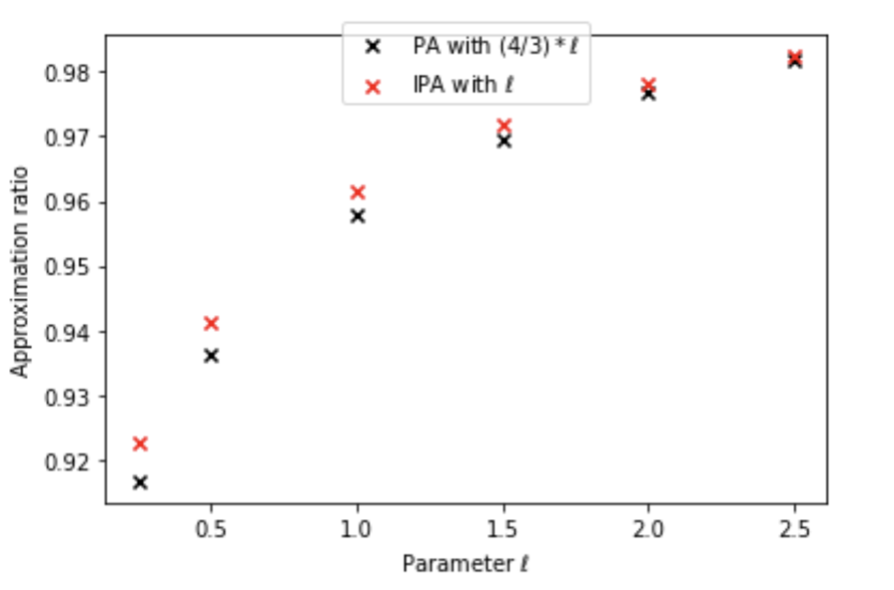}
    \caption{Approximation ratio} 
    \end{subfigure}
    \caption{Comparison of IPA$(\ell)$ and PA$((4/3)\ell)$: fairness profiles for October (left) and social welfare on the full dataset (right).}
    \label{fig:IPAPAsocialwelfare}
\end{figure*}

\paragraph{Evaluating bid-stability.}
The main difference between our theoretical model and the data in practice is that \textit{not every advertiser bids on every keyword}. In fact, only a small number of advertisers bid on each keyword. 
We thus need to modify how we measure the similarity between keywords. For a keyword $u$, let $S_u$ denote the set of advertisers who place a nonzero bid on $u$. We make two changes. First, we only consider pairs of keywords $u$ and $v$ that share enough advertisers, by placing a lower bound on the Jaccard similarity between $S_u$ and $S_v$. Specifically, we consider pairs $u$ and $v$ with $J(S_{u}, S_{v}) := \frac{|S_{u} \cap S_{v}|}{|S_{u} \cup S_{v}|}\ge 0.67$. Second, we measure bid similarity as well as allocation difference for $u$ and $v$ only over the advertisers in $S_{u} \cap S_{v}$. 
Formally, for keywords $u$ and $v$ with bid vectors $\bid$ and $\bid'$, we define:
\begin{align*}
    & \tilde{d}(u, v) := \max_{i \in S_{u} \cap S_{v}}|\alloci(\bid) - \alloci(\bid')|,  \\
    & \tilde{\stabpar}(u, v) := \max_{i \in S_{u} \cap S_{v}} \max\left(\frac{\bidi}{\bidi'}, \frac{\bidi'}{\bidi} \right).
\end{align*}
The bid stability achieved by an algorithm is captured by the set of $(\tilde{\stabpar}(u, v), \tilde{d}(u, v))$ across all pairs $(u, v)$ where $J(S_{u}, S_{v}) \ge 0.67$. In order to provide a summary of these guarantees, we build the following \textit{bid stability profile}. We consider buckets of width $0.1$ in $[1,2]$. For each bucket $B$, we consider the set $\left\{ \tilde{d}(u, v) \mid \tilde{\stabpar}(u, v) \in B\right\}$, and estimate the 90th percentile allocation difference over this set by sampling pairs from the bucket. In other words, the bid-stability profile maps each similarity bucket to the allocation difference guarantee achieved by at least 90\% of the pairs of keywords in that bucket. 

\paragraph{Highest-bid-wins auction is not bid-stable.} Our first finding is that the highest-bid-wins auction is not bid stable. Recall that for each keyword this auction places the full allocation on the advertiser with the highest bid. We observe that for many pairs of highly similar keywords the highest bidder turns out to be different, leading to unfair allocation. Figure \ref{fig:highestbid} shows the fraction of pairs within each bucket with an allocation difference of $1$. Across all months and across all buckets, the highest-bid-wins auction uniformly has at least 30\% of allocation differences at $1$. This demonstrates that the highest-bid-wins auction is indeed fundamentally incompatible with
bid stability.

\vspace{0.1in}
\paragraph{Analysis of IPA.} Unlike the highest-bid-wins auction, IPA achieves strong bid stability on the Yahoo A1 dataset, thus aligning with our theoretical findings from Section \ref{sec:inverseprop}. Figure~\ref{fig:theory}(c) shows the bid stability profile for IPA in the month of July across different parameter settings of $\ell$. (Similar patterns can be observed on data from other months.) All of the IPA parameterizations achieve strong bid stability guarantees, exhibiting significant improvement over theoretical bounds. E.g., the 90th percentile allocation difference is close to $0$ at $\tilde{\stabpar}=1.0$ and never exceeds $0.15$. As predicted by Theorem \ref{thm:fairness}, the bid stability guarantees weaken as $\ell$ increases. 

In addition to bid stability, IPA achieves close-to-optimal social welfare. In Figure~\ref{fig:IPAPAsocialwelfare}(b), we show the approximation ratio achieved by IPA over the entire dataset. IPA with $\ell=2.5$ achieves over $98\%$ of the optimal social welfare. As predicted by Theorem \ref{thm:fairvalue}, the approximation ratio worsens as $\ell$ decreases, but even at $\ell=0.1$, IPA obtains over $92\%$ of the optimum. Figure~\ref{fig:theory}(d) provides a breakdown of IPA's performance on different slices of the dataset, with each slice corresponding to keywords that receive a specific number of nonzero bids from distinct advertisers. Note that IPA's approximation ratio does not change much with $k$, especially when $\ell$ is sufficiently high. This aligns with our theoretical bounds. Finally, we note that IPA consistently outperforms the theoretical bounds from Theorem \ref{thm:fairvalue}.\footnote{Our worst-case bounds are tight for settings where one advertiser places a far higher bid than all others. It is likely that bid vectors with this structure occur infrequently in practice.}

\vspace{0.1in}
\paragraph{Comparison of IPA and PA.} 
First, we describe how we select parameters for both IPA and PA in order to perform a proper comparison of the two algorithms. This is a delicate task because both the \textit{shape} and \textit{magnitude} of the bid stability profiles can vary across the different algorithms. Recall that the bid stability profile maps each similarity bucket to a 90th percentile allocation difference. Let $F_1(\tilde{\stabpar})$ and $F_2(\tilde{\stabpar})$ denote these 90th percentiles for the two algorithms for the bucket corresponding to bid similarity $\tilde{\stabpar}$. We consider the ratio $F_1(\tilde{\stabpar})/F_2(\tilde{\stabpar})$ and want it to be as close to $1$ as possible. Accordingly, for every $\ell$ and $\ell'$, we consider the spread of this ratio across all buckets, and pick the $\ell'$ for which this spread is as small as possible while being close to $1$. We find that $\ell'=\frac 43\ell$ provides the best match. See, e.g., Figure~\ref{fig:IPAPAsocialwelfare}(a), which corresponds to the month of October.\footnote{In Section \ref{sec:inverseprop}, we instead compared IPA with parameter $\ell$ and PA with parameter $2 \ell$. However, on this dataset, the ratio falls below $0.5$ in some months, and so IPA with $\ell$ and PA with $2\ell$ do not achieve comparable bid-stability guarantees.} The ratio turns out to lie in $[0.6, 1.35]$ for every month, although the actual values vary substantially across months.

Figure \ref{fig:IPAPAsocialwelfare} shows the approximation ratios for social welfare of IPA($\ell$) and PA($\frac 43\ell$). Note that across all values of $\ell$, IPA consistently outperforms PA with respect to social welfare. Figure~\ref{fig:theory}(d) shows the worst case approximation ratio for social welfare of IPA($\ell$) versus PA($\frac 43\ell$) on slices of the dataset consisting of specific numbers of advertisers. Although PA outperforms IPA for 2 advertisers, the performance of PA quickly degrades as the number of advertisers increases. On the other hand, the performance of IPA stays more or less constant across different values of $k$. This aligns with our theoretical findings. 
Altogether we conclude that IPA outperforms PA for most parameter settings.

\section{Subset fairness}\label{sec:subset}
\newcommand{\sets}{{\mathcal C}}
\newcommand{\w}{{\omega}}
\newcommand{\cw}{{\omega_{\text{cluster}}}}
\newcommand{\pw}{{\omega_{\text{part}}}}

By satisfying value stability, the mechanisms that we design guarantee that if two users have similar values from all advertisers, then the difference between the allocation that they receive from any advertiser must be small. However, in \cite{CIJ20}, we observed that in some settings this notion is too weak. To take an example, consider job ads, and suppose that two users Alice and Bob have similar value vectors across job advertisers, but that all high paying job ads value Alice slightly less than Bob. Then, a value-stable ad auction may show every high paying job ad with slightly less probability to Alice than to Bob. However, considering the set of all high paying jobs together, Alice may see a job in this set with far lower probability than Bob. The approach in \cite{CIJ20} is to impose value-stability with respect to arbitrary such groups of advertisers.\footnote{In \cite{CIJ20}, this notion is referred to as total-variation fairness. This corresponds to ensuring that the distributions over ads assigned to similar users are not only close under $\ell_\infty$ distance, but also close under $\ell_1$ or total variation distance.} This guarantees that users are not discriminated against even when considering sets of related ads rather than just a single ad.

Our family of IPA auctions does not always satisfy value-stability with respect to arbitrary such groups of advertisers (as we discuss in Section \ref{subsec:totalvargap}). Indeed it appears to be challenging to satisfy this stronger fairness property while also guaranteeing a constant factor approximation in social welfare. In this work, we take a more nuanced approach and consider notions that interpolate between value-stability on each advertiser and value-stability on all possible groups of advertisers. Specifically, we require the algorithm to satisfy value-stability with respect to a restricted collection of sets over advertisers. 

\begin{definition}
\label{def:subsetvaluestable}
Let $\mathcal{C}$ be a collection of subsets of $[k]$.
An allocation is \textbf{\SubsetValueStable} with respect to $\mathcal{C}$ and a function $f: [1, \infty] \rightarrow [0,1]$ if the following condition is satisfied for every pair of value vectors $\val$ and $\val'$ and for every $C \in \mathcal{C}$:
\[\left|\sum_{i \in C} \alloci(\val) - \sum_{i \in C} \alloci(\val')\right| \le f(\stabpar) \text{ for all } i \in [k], \,\, \text{where }\stabpar \text{ is defined as } \max_{i\in [k]} \left( \max\left\{\frac \vali{\vali'}, \frac {\vali'}\vali\right\}\right).\]

\end{definition}

We envision that a trusted authority selects the collection $\mathcal{C}$ over which value-stability must be guaranteed. This may include, for example, all job ads belonging to a particular category or a particular compensation bracket, etc. When $\mathcal{C}$ contains all subsets of $[k]$, then this definition provides the guarantees proposed in \cite{CIJ20}. When $\mathcal{C}$ only consists of subsets of size $1$, then this definition reduces to the usual definition of value-stability. For intermediate collections $\mathcal{C}$, the definition is stronger than the usual definition of value-stability but not as strong as the notion proposed in \cite{CIJ20}. For convenience, we often do not include singleton sets by default in $\mathcal C$. Our intention is for the algorithm to satisfy \SubsetValueStability in addition to value-stability.

In Section \ref{subsec:totalvargap}, we first argue that \SubsetValueStability comes at a cost: the \InvProp allocation algorithm defined in Section~\ref{sec:inverseprop} satisfies value-stability, but we show that it does not satisfy \SubsetValueStability for $\mathcal{C} = 2^{[k]}$. In fact, there is a performance gap between algorithms that satisfy \SubsetValueStability for $\mathcal{C} = 2^{[k]}$ and those that only satisfy value-stability. In Section \ref{subsec:subsetfairness}, we develop auctions that guarantee \SubsetValueStability for any given collection $\mathcal{C}$, while obtaining approximation ratios that degrade with the complexity of $\mathcal{C}$.

\subsection{A gap between value-stability and \SubsetValueStability}\label{subsec:totalvargap}
We consider the strongest version of \SubsetValueStability, where $\mathcal{C} = 2^{[k]}$. 
We first show that the class of inverse-proportional allocation algorithms described in Section~\ref{sec:inverseprop} does not satisfy \SubsetValueStability with respect to $\mathcal{C} = 2^{[k]}$. 

\begin{example}
\label{example:tvbroken}
Let $k$ be any even number and let $\stabpar > 1$ be a parameter that we will specify later. Let $\val$ and $\val'$ be defined as follows: suppose that $\vali[j] = 1$ for $1 \le j \le k/2$ and  $\vali[j] = \stabpar$ for $k/2 +1 \le j \le k$;  whereas $\valprimei[j] = \stabpar$ for $1 \le j \le k/2$ and  $\valprimei[j] = 1$ for $k/2 +1 \le j \le k$. 

We now observe that for some (not too small) value of $\stabpar$, the inverse-proportional allocation algorithm with parameter $\ell$ will assign $\allocprimei[i] = 0$ for all $1 \le i \le k/2$ and $\alloci[j] = 0$ for all $k/2 + 1 \le j \le k$. As a result, it would hold that $\sum_{i=1}^{k/2-1} \alloci[i] = 1$ and $\sum_{i=1}^{k/2-1} \allocprimei[i] = 0$. In fact it suffices to set $\stabpar= \left(\frac{k/2-1}{k/2}\right)^{1/\ell}$  so that $f_{\ell}(\stabpar) = 1 - \left(\frac{k/2 - 1}{k/2}\right)^2\approx 4/k$.\footnote{With these values, we have $(k-1)\frac{(k/2)/(k/2-1)}{k/2 + (k/2)((k/2)/(k/2-1))} = 1$, and so the advertisers with values $\stabpar$ will get dropped.}
\end{example}

Example \ref{example:tvbroken} shows that even for two value vectors that are very close to each other, there may exist a subset of advertisers on which one value vector receives an allocation of $1$ while the other value vector receives an allocation of $0$. Thus, the example violates \SubsetValueStability with respect to $\mathcal{C} = 2^{[k]}$ and $f_{\ell}$. Is it possible to construct a different allocation algorithm that is \SubsetValueStable for $\mathcal{C}$  but performs as well as \InvProp allocation on social welfare? We show that this is not possible for small values of $\ell$: there is a gap between the competitive ratio achievable by any total-variation fair algorithm and the competitive ratio of the \InvProp algorithm.
 
\begin{theorem}
\label{thm:gap}
For any $0 < \ell < \infty$, let $\ubTV(k,\ell)$ be the optimal competitive ratio achievable by any prior-free, anonymous allocation algorithm that achieves \SubsetValueStability for $\mathcal{C} = 2^{[k]}$ and $f_{\ell}$. Let $\lb(\ell)$ denote the competitive ratio achieved by the \InvProp allocation algorithm with parameter $\ell$. Then, it holds that $\limsup_{k \rightarrow \infty} \ubTV(k,\ell) \le \frac{2\ell}{2\ell+1}$ and $\liminf_{k \rightarrow \infty} \frac{\lb(\ell)}{\ubTV(k,\ell)} \ge  \frac{2 \ell+1}{\ell + 1} \left(\frac{1}{2} + \frac{1}{2 \ell} \left(1 - \frac{\ell^\ell}{(\ell+1)^\ell} \right) \right)$. In particular, $\liminf_{\ell \rightarrow 0} \left(\liminf_{k \rightarrow \infty} \frac{\lb(\ell)}{\ubTV(k,\ell)}\right) = \infty$. 
\end{theorem}

\subsection{\CapitalSubsetValueStability over restricted set collections}\label{subsec:subsetfairness}
Given the gap in Theorem~\ref{thm:gap}, we ask whether good competitive ratios can be achieved if \SubsetValueStability is only required over some restricted collection of sets of advertisers as opposed to arbitrary subsets of advertisers. We show in the remainder of this section that such improved guarantees are indeed possible.  In this section, we describe and motivate three kinds of set collections that enable positive results.

As a warm-up, one might ask whether the challenge with ensuring \SubsetValueStability for $\mathcal{C} = 2^{[k]}$ is that there are exponentially many fairness constraints to satisfy -- one for each possible subset of advertisers. This turns out to not be the case. Indeed, the proof of Theorem \ref{thm:gap} relies on requiring \SubsetValueStability with respect to a collection of only $\Theta(\log k)$ subsets of advertisers. 

Our first observation is that \SubsetValueStability over sets of small size can be achieved without much loss in performance. To formalize this, we define the ``bandwidth'' of a collection of sets---the size of the largest set in the collection---as a measure of complexity of the set system:
\begin{definition}
\label{def:width}
  The {\em width} of a collection $\sets\subset 2^{[k]}$ of subsets of advertisers, denoted $\w$, is defined as $\max_{C\in\sets} |C|$.
\end{definition}

While collections of small sets may not be very interesting in themselves, this definition leads us to a more nuanced notion of complexity of set systems. Given a set system $\sets\subset 2^{[k]}$, we will say that two advertisers $i$ and $j$ are equivalent if they belong to exactly the same sets in $\sets$: for all $C\in\sets$, $i\in C \iff j\in C$. This partitions advertisers into equivalence classes or clusters. Let $L_1, L_2, \cdots$ denote these clusters. Every set $C\in\sets$ is then the union of some subset of the clusters. We will define the ``cluster bandwidth'' of a collection of sets, denoted $\cw$, as the maximum over all sets $C\in\sets$ of the number of clusters contained in $C$.
\begin{definition}
\label{def:clusterwidth}
  Given the partition $\left\{L_1, L_2, \cdots\right\}$ of advertisers into clusters, as described above, the {\em cluster bandwidth} of a collection $\sets\subset 2^{[k]}$ of subsets of advertisers, denoted $\cw(C)$, is defined as $\max_{C\in\sets} |\{i:L_i\subset C\}|$.
\end{definition}

Many interesting set collections can have low cluster bandwidth. For example, when the sets $C\in\sets$ are disjoint, the cluster bandwidth of the collection is simply $1$. Likewise, when advertisers have few relevant attributes, the cluster bandwidth is no larger than the number of different values the attributes can take. For example, suppose that we classify job ads according to whether they are high pay, medium pay, or low pay jobs, and whether they are tech sector, or finance, or academic jobs. Then there are nine possible kinds of ads and the sets in our collection may correspond to some subset of these nine types. Then, the cluster bandwidth of this set collection is no more than nine.

Finally, we consider settings where the \SubsetValueStability constraints apply only over small sub-categories of advertisers. For example, we may be interested in providing fairness guarantees across arbitrary sets of job ads relevant to a single geographical area. In this case, we would partition the set of all advertisers according to geographic location, and then enforce \SubsetValueStability over arbitrary subsets that lie entirely within a single component of the partition. The performance of our allocation algorithm will then depend on the sizes of the components of the partition.
\begin{definition}
  Let $\{L_1, L_2, \cdots\}$ be a partition of advertisers into clusters. Let $\sets\subset 2^{[k]}$ be a set collection where for each set $C\in\sets$, there exists an index $i$ with $C\subset L_i$. We define the {\em partitioned width} of $\sets$, denoted $\pw(C)$, as $\max_i |L_i|$.
\end{definition}

\subsubsection{Algorithms achieving \SubsetValueStability}\label{subsubsec:algorithmsset}

We now construct variants of \InvProp allocation that are competitive for social welfare while satisfying \SubsetValueStability over small-width set collections as defined in the previous subsection. We use three building blocks for our constructions, put together in novel ways: the \InvProp allocation algorithm, the capped inverse-proportional allocation algorithm (Algorithm \ref{alg:cappedIPA}) with different settings of $\multpar < 1$, and the proportional allocation algorithm of \cite{CIJ20}. As described previously, \InvProp allocation aggressively favors high values but hurts advertisers with low values when these are numerous. Capped inverse-proportional allocation places an upper bound on the maximum allocation received by any advertiser and thus turns out to provide stronger \SubsetValueStability guarantees while maintaining a good competitive ratio. At the other extreme, proportional allocation provides very strong \SubsetValueStability guarantees, but the competitive ratio suffers. Thus, the three types of allocation rules address different aspects of \SubsetValueStable allocation. However, putting them requires care so as to not magnify differences in allocation across different users.

We first describe how capped inverse-proportional allocation (Algorithm \ref{alg:cappedIPA}) achieves 
\SubsetValueStability across small sets while maintaining a good competitive ratio. We already showed in Corollary \ref{cor:cappedIPA} that capped inverse-proportional allocation with parameters $\ell$ and $1/\sizepar$ achieves at least a $\alpha_{\ell}/\sizepar$ competitive ratio; now, we show that this algorithm satisfies \SubsetValueStability on all subsets of size $\sizepar$.
\begin{theorem}
\label{thm:case3}
Let $\mathcal{C}$ be a collection such that $\w(\mathcal{C}) \le \sizepar$. For any $0 < \ell < \infty$, the capped inverse-proportional allocation algorithm with parameters $g(x) = x^{-\ell}$ and $1/\sizepar$ achieves \SubsetValueStability with respect to $\mathcal{C}$ and $f_{\ell}(\stabpar) = 1 - \stabpar^{-2\ell}$.
\end{theorem}
\begin{proof}
Consider two value vectors $\val$ and $\valprime$ such that $\max_{i\in [k]} \left(\max\left\{\frac \vali{\vali'}, \frac {\vali'}\vali\right\}\right) = \stabpar$. For value-stability, by Theorem \ref{thm:fairness}, we know that the inverse-proportional allocation algorithm with parameter $g(x) = x^{-\ell}$ assigns allocations $\alloci[i]$ and $\allocprimei[i]$ that satisfy $|\alloci[i] - \allocprimei[i]| \le f(\stabpar)$. Now, notice that the capped inverse-proportional allocation algorithm assigns allocations $\smallalloci[i] = \frac{\alloci[i]}{\sizepar} + \frac{1}{k} \left(1 - \frac{1}{\sizepar}\right)$ and $\smallallocprimei[i] = \frac{\allocprimei[i]}{\sizepar} + \frac{1}{k}\left(1 - \frac{1}{\sizepar}\right)$. We see that $|\smallalloci[i] - \smallallocprimei[i] | = \frac{1}{\sizepar} |\alloci[i] - \allocprimei[i]| \le \frac{1}{\sizepar} f(\stabpar)$. Now, let $C$ be a subset in $\mathcal{C}$. We know that $|C| \le \sizepar$, so $\left|\sum_{i \in C} \smallalloci[i] - \sum_{i \in C} \smallallocprimei[i]\right| \le \sum_{i \in C} \left|\smallalloci[i] - \smallallocprimei[i]\right| \le |C| f(\stabpar) / \sizepar \le f(\stabpar)$.  
\end{proof}

Next we combine capped inverse proportional allocation with inverse proportional allocation to handle set collections with small cluster width. At a high level, we use the capped inverse-proportional allocation algorithm  to first allocate probabilities to each cluster of advertisers as defined in Definition~\ref{def:clusterwidth}, and then use the inverse-proportional allocation algorithm to further subdivide the allocation to advertisers within each cluster. More formally:
\begin{alg}
\label{alg:intersections}
The algorithm with parameters $\ell$ and $\sizepar$ operates as follows:
\begin{enumerate}
\item Partition advertisers into clusters $\mathcal{L} = \left\{L_1, L_2, \cdots \right\}$, as defined in Definition~\ref{def:clusterwidth}.
\item Use the capped inverse-proportional allocation algorithm (Algorithm \ref{alg:cappedIPA}) with parameters $g(x) = x^{-\ell}$ and $1/\sizepar$ on $\left\{L_1, L_2, \cdots \right\}$, treating each set $L_i$ as an advertiser with value $\max_{j \in L_i} \vali[j]$. Obtain an allocation $\left\{\bigalloci[L_i]\right\}_{L_i \in \mathcal{L}}$. 
\item For each $L_i \in \mathcal{L}$, run the inverse-proportional allocation algorithm with parameter $g(x) = x^{-\ell}$ to obtain an allocation $\left\{\smallalloci[j]\right\}_{j \in L_i}$. Now, set $\alloci[j] = \smallalloci[j] \cdot \bigalloci[L_i]$. 
\end{enumerate}
\end{alg}
This algorithm satisfies nice properties in terms of value-stability and competitive ratio:
\begin{theorem}
\label{thm:case2}
Consider collection of subsets $\mathcal{C}$ such that $\cw(\mathcal{C}) \le n$. Then, Algorithm \ref{alg:intersections} with parameters $\ell$ and $\sizepar$ satisfies \SubsetValueStability with respect to $\mathcal{C}$ and $f_{\ell}(\stabpar) = 1 - \stabpar^{-2\ell}$ as well as value-stability for $f(\stabpar) = 2 \cdot f_{\ell}(\stabpar) = 2(1 - \stabpar^{-2\ell})$. Furthermore, the algorithm achieves a competitive ratio of at least $\frac{1}{n} \alpha^2_{\ell}$. 
\end{theorem}
\begin{proof}
Consider two value vectors $\val$ and $\valprime$ such that $\max_{i\in [k]} \left(\max\left\{\frac \vali{\vali'}, \frac {\vali'}\vali\right\}\right) = \stabpar$.
To show \SubsetValueStability we use that every set in $\mathcal{C}$ is a union of at most $\sizepar$ sets in $\mathcal{L}$. Suppose that $C = L_{i_1} \cup L_{i_2} \ldots \cup L_{i_{\sizeparprime}}$ where $\sizeparprime \le \sizepar$.  
Using Theorem \ref{thm:case3}, we know that:
\[\left| \sum_{j \in C} \alloci[j] - \sum_{j \in C} \allocprimei[j] \right| = \left|\sum_{l=1}^\sizeparprime \bigalloci[L_{i_l}]- \sum_{l=1}^\sizeparprime \bigallocprimei[L_{i_l}] \right| \le 1 - \stabpar^{-2\ell}.\]
To show value-stability, we use the fact that for any $L_i \in \mathcal{L}$, we know $|\bigalloci[L_i] - \bigallocprimei[L_i]| \le 1 - \stabpar^{-2\ell}$ by Theorem \ref{thm:case3}, and for any $j \in L_i$, we know $|\smallalloci[j] - \smallallocprimei[j]| \le 1 - \stabpar^{-2\ell}$ by Theorem \ref{thm:fairness}. Thus, we know that $|\alloci[j] - \allocprimei[j]| = |\smallalloci[j] \cdot \bigalloci[L_i] - \smallallocprimei[j] \bigalloci[L_i]| \le |\smallalloci[j] - \smallallocprimei[j]|\bigalloci[L_i] + |\bigalloci[L_i] - \bigallocprimei[L_i]| \smallallocprimei[j] \le 2 (1 - \stabpar^{-2\ell})$. The approximation ratio follows from Corollary \ref{cor:cappedIPA} (on the first-level allocation) and Theorem \ref{thm:fairvalue} (on the second-level allocation). 
\end{proof}

Finally, we show that a similar two-step composition provides fairness for set collections with small partitioned width. For this setting, we compose the inverse-proportional allocation algorithm with the proportional allocation algorithm in \cite{CIJ20}. In particular, denoting the partition over advertisers as $L_1, L_2, \cdots$, we use Algorithm \ref{alg:cappedIPA} to perform the allocation across different $L_i$s, and then use the proportional allocation algorithm to divide the allocation within each $L_i$. More formally:
\begin{alg}
\label{alg:hierarchy}
The algorithm with parameters $\ell$ and $\mathcal{L} = (L_1, L_2 \ldots)$ operates as follows:
\begin{enumerate}
\item Use the inverse-proportional allocation algorithm with parameter $g(x) = x^{-\ell}$ on $(L_1, L_2 \ldots)$, treating each set $L_i$ as an advertiser with bid $\max_{j \in L_i} \vali[j]$. Obtain an allocation $\left\{\bigalloci[L_i]\right\}_{L_i \in \mathcal{L}}$. 
\item For each $L_i \in \mathcal{L}$, run the proportional allocation algorithm with parameter $g(x) = x^{2 \ell}$ to obtain an allocation $\left\{\smallalloci[j]\right\}_{j \in L_i}$. Now, set $\alloci[j] = \smallalloci[j] \cdot \bigalloci[L_i]$. 
\end{enumerate}
\end{alg}
This algorithm achieves \SubsetValueStability over the collection $\sets=\cup_i 2^{L_i}$ and competitive ratio guarantees that degrade with $\pw(\sets)$.  
\begin{theorem}
\label{thm:case1} Let $\mathcal{L} = (L_1, L_2, \ldots)$ be a partition of $[k]$. Consider a collection $\sets=\cup_i 2^{L_i}$ such that $\pw(\sets)= \sizepar$. Algorithm \ref{alg:hierarchy} with parameters $\ell$ and $\mathcal{L}$ satisfies \SubsetValueStability with respect to $\sets$ and $f(\stabpar) = 2 \cdot f_{\ell}(\stabpar) =  2(1 - \stabpar^{-2\ell})$ as well as value-stability with respect to $\sets$ and $f(\stabpar) = 2 \cdot f_{\ell} = 2(1 - \stabpar^{-2\ell})$. Moreover, the algorithm achieves a competitive ratio of at least $\sizepar^{-\frac{2}{\ell}} \alpha_{\ell}$. 
\end{theorem}
\begin{proof}
Consider two value vector $\val$ and $\valprime$ such that $\max_{i\in [k]} \left(\max\left\{\frac \vali{\vali'}, \frac {\vali'}\vali\right\}\right) = \stabpar$.
We first show \SubsetValueStability and value-stability. We use the fact that for any $L_i \in \mathcal{L}$, we know $\left|\bigalloci[L_i] - \bigallocprimei[L_i]\right| \le f_{\ell}(\stabpar)$ by Theorem \ref{thm:fairness}. Moreover, for any $L_i \in \mathcal{L}$ and for any $S \subseteq L_i$, we know $\left|\smallalloci[j] - \smallallocprimei[j]\right| \le f_{\ell}(\stabpar)$ by the properties of the proportional allocation algorithm shown in \cite{CIJ20}. Thus, for every $S \subseteq L_i$, we know that $\left|\sum_{j \in S} \alloci[j] - \sum_{j \in S} \allocprimei[j]\right| = \left|\left(\sum_{j \in S} \smallalloci[j] \right) \cdot \bigalloci[j]  - \left(\sum_{j \in S} \smallallocprimei[j] \right) \bigallocprimei[L_i]\right|\le \left|\sum_{j \in S} \smallalloci[j] - \sum_{j \in S} \smallallocprimei[j]\right| \bigalloci[L_i] + |\bigalloci[L_i] - \bigallocprimei[L_i]| \left(\sum_{j \in S} \smallallocprimei[j]\right) \le 2 f_{\ell}(\stabpar)$. The competitive ratio follows from Theorem \ref{thm:fairvalue} (on the first-level allocation) and the competitive ratio bound in \cite{CIJ20} (on the second-level allocation). 
\end{proof}

\section{Future work}\label{sec:futurework}

In this work we address the tradeoffs between fairness and social welfare through a simplistic model that assumes that advertisers bid to maximize their utility in every auction. In reality, advertisers have budgets and optimize for their long term returns. Incorporating fairness constraints into these more general models of ad auctions is an interesting avenue for future work. For the model we study, our work presents an alternative to the proportional allocation mechanism of \cite{CIJ20}. The two classes of allocation algorithms perform well under different contexts, and it would be interesting to understand whether one can interpolate between the two to obtain fairness guarantees as strong as those of proportional allocation while at the same time achieving performance commensurate with that of inverse proportional allocation. Finally, a very interesting open direction is to examine ad auction design under a group fairness constraint.

\bibliography{bibliography.bib}

\newpage 

\appendix

\section{Multi-Category Setting}\label{appendix:multicategory} 
Platforms often service advertisers from many different categories (e.g. jobs advertisers, credit advertisers, housing advertisers, etc.) that compete against each other for the same users. The challenge is that for different categories, what constitutes similarity between users might be very different. For example, two users who are considered similar in the context of credit ads might be considered very different in the context of job ads. In \cite{CIJ20}, we showed that in this setting, because of differing amounts of competition for different users, no mechanism can simultaneously achieve multiple-task fairness as well as good social welfare. In fact, we construct examples where multiple-task fairness results in allocations that are worse for all users than certain allocations that do not satisfy multiple-task fairness. In this context, we advocated for instead requiring envy-freeness across categories. We proposed a new notion of \textit{compositional fairness} where every user is allowed to select a few categories of their choice with the dual guarantee that:
\begin{enumerate}
    \item The user's total allocation of ads within their favored categories is as large as that of any other user (envy-freeness).
    \item The mix of ads that the users receive within any category is fair according to the single-category fairness definition discussed previously (value-stability). 
\end{enumerate}
The allocation algorithms we develop in this work for a single category of ads combine in a straightforward fashion with \cite{CIJ20}'s envy-free algorithms for across-category fairness to produce allocations that satisfy compositional fairness. In doing so, we significantly improve upon the approximation ratios in the previous work, since our guarantees from within-category allocations carry over to the compositional setting. 

\section{Payment Rule}\label{sec:payment} 
For the case of two advertisers, $k=2$, supporting payments for the IPA can be calculated easily. Here is an explicit formula for the payment of advertiser $1$ when the algorithm is parameterized by the function $g(x)=1/x$.
\[\payi[1] = \vali[2] \ln\left(1 + \frac{\vali[1]}{\vali[2]}\right) - \frac{\vali[1]\vali[2]}{\vali[1] + \vali[2]}. \]

Now, we consider $k > 2$. Here is a formula for the payment of advertiser $1$ when the algorithm is parameterized by the function $g(x)=1/x$. If $\alloci[1] = 0$, then $\payi[1] = 0$. Otherwise, $\payi[1]$ has the following form. WLOG, suppose that $\vali[2] \le \ldots \le \vali[k]$. We know that the set of advertisers who receive nonzero allocation can be expressed as $\left\{1 \right\} \cup [m, k]$ for some $2 \le m \le k$. Furthermore, let $[m', k]$ be the set of advertisers who receive nonzero allocation on $[0, \vali[2], \ldots, \vali[k]]$. Let $c_{m'-1} = (\sum_{j=m'}^k \vali[j]^{-1})/(k-m')$, let $c_i = (k-i)\vali[i]^{-1}- \sum_{j=i+1}^k v_j^{-1}$ for $m' \le i \le m - 1$, and let $c_m = 1/\vali[1]$. Then the payment is equal to: 
\[\payi[1] = \frac{k-m'}{\sum_{j=m'}^k \vali[j]^{-1}} - \frac{k-m+1}{\vali[1]^{-1} + \sum_{j=m}^k \vali[j]^{-1}} +  \sum_{i=m'}^{m} \left(\frac{k-i+1}{\sum_{j=i}^k \vali[j]^{-1}} \log\left(\frac{1 + (1/c_i) \sum_{j=i}^k \vali[j]^{-1}}{1 + (1/c_{i-1}) \sum_{j=i}^k \vali[j]^{-1}} \right) \right).\]  

\section{Auxiliary Propositions and Proofs}\label{sec:proofs}
We prove an auxiliary proposition used in the proof of Theorem \ref{thm:fairness}.
\begin{proposition}
\label{prop:derivative}
Let $c < 1$ be a constant. Consider the function $h(x) = \frac{1}{1 + x c} - \frac{1}{1 + x}$ on $x \ge 0$. Thus function is increasing when $x \le \frac{1}{\sqrt{c}}$ and decreasing when $x \ge \frac{1}{\sqrt{c}}$. Thus, the maximum is attained when $x = \frac{1}{\sqrt{c}}$. (When $c = 1$, $h(x)$ is always $0$.)
\end{proposition}
\begin{proof}
The derivative with respect to $x$ is $-\frac{c}{(1 + x c)^2} + \frac{1}{(1+x)^2}.$ The function is weakly decreasing if and only if  $\frac{c}{(1 + x c)^2} \ge \frac{1}{(1+x)^2}$. This is equivalent to $c + 2xc + x^2 c \ge 1 + x^2 c^2 + 2 x c$. Rearranging yields $c - 1 \ge x^2 c (c - 1)$. Since $c < 1$, this is equivalently $1 - c \le x^2 c(1 - c)$, which is equivalently $x \ge \frac{1}{\sqrt{c}}$.   
\end{proof}

We prove an auxiliary proposition used in the proof of Theorem \ref{thm:nearoptimality}. 
\begin{proposition}
\label{prop:boundinglogs}
Let $0 < c < 1$ and $r > 0$ be parameters. For any $x \in [1, \infty)$, it holds that: 
\[\min(c, \ln(x) \cdot r c) \ge c(1 - x^{-r}).\]
\end{proposition}
\begin{proof}
We see that $c(1 - x^{-r}) \le c$ trivially. Thus, it suffices to show that $c(1 - x^{-r}) \le \ln(x) \cdot r c$. Rearranging, this condition can be written as $1 - x^{-r} \le \ln(x^r)$, which can be written as $x^{-r} - 1 \ge \ln(x^{-r})$. Now, the inequality follows from the fact that $\ln(1 + y) \le y$. 
\end{proof}

We now prove Theorem \ref{thm:gap}. In the proof of Theorem \ref{thm:gap}, we use the following construction of value vectors.

\begin{example}
\label{example:gap}
Let $0 < \gamma < 1$, $0 < \ell < \infty$, and $n \in \mathbb{N}$ be parameters. We construct an instance of $k = 2^n$ advertisers and $n+1$ value vectors as follows. All of the value vectors will be permutations of the same multi-set of values, but the ordering across advertisers will differ. We define the multi-set of values $V$ as follows: $V$ will have $1$ copy of $1$, and for $1 \le i \le n$, $V$ will have $2^{i-1}$ copies of $\gamma^i$. 

We now design a sequence of $n+1$ value vectors $\valnum[0]$, $\valnum[1]$, $\valnum[2]$, $\ldots$, $\valnum[n]$. The value vector $\valnum[0]$ will organize the values in $V$ so that $\valnumi{0}{1} \ge \valnumi{0}{2} \ge \ldots \ge \valnumi{0}{k}$. For $1 \le i \le n$, the value vector $\valnum[i]$ will organize the values so that $\valnumi{i}{j} = \valnumi{0}{j}$ for $2^i + 1 \le j \le k$, and $\valnumi{i}{j} = \valnumi{0}{2^i+1-j}$ for $1 \le j \le 2^i$ (i.e. the first $2^{i-1}$ values are flipped with the next set of $2^{i-1}$ values, and everything else stays the same.) 
\end{example}

\begin{proof}[Proof of Theorem \ref{thm:gap}]
To compute an upper bound on $\ubTV(k, \ell)$, we
use the construction in Example \ref{example:gap}. The fact that the algorithm is prior-free and anonymous requires that the allocation assigned to each of these value vectors is always the same up to the appropriate permutation. For each $0 \le i \le n$, suppose that the algorithm places a \textit{total allocation} of $\alloci[i]$ on the $2^i$ copies of $\gamma^i$. We know that the approximation ratio is thus $\sum_{i=0}^n (\alloci[i] \cdot \gamma^i)$. 
 
Now, we require \SubsetValueStability on the collection of sets $S_{i} := \left\{1, \ldots, 2^{i-1}\right\}$ for $1 \le i \le n$. Notice that this collection only has $\Theta(n) = \Theta(\log k)$ sets. In particular, we enforce the constraint for $S_i$ on value vectors $\valnum[0]$ and $\valnum[i]$. For $\valnum[0]$ there is a total allocation of $\alloci[0] + \ldots + \alloci[i-1]$ on $S_i$; for $\valnum[i]$ there is a total allocation of $\alloci[i]$ on $S_i$. This means that: 
\[\alloci[0] + \ldots + \alloci[i-1] - \alloci[i] \le f(\gamma^{-i}). \] The other constraint is that $\alloci[0] + \ldots + \alloci[n] \le 1$ since the total allocation cannot exceed 1. It is easy to see that the social welfare is thus maximized when all of the inequalities are tight. This means that $\alloci[0] = \frac{1}{2^n} + \sum_{i=1}^n \frac{f(\gamma^{-i})}{2^i}$ and for $1 \le j \le n$, $\alloci[j] = \frac{1}{2^{n-j+1}} + \left(\sum_{i=j+1}^n \frac{f(\gamma^{-i})}{2^{i-j+1}} \right) - \frac{f(\gamma^{-j})}{2}$. Observe that this allocation is always nonnegative (and thus valid) because $f$ is an increasing function. 

This means that:
\[\ubTV(k, \ell) \le \left(\frac{1}{2^n} + \sum_{i=1}^n \frac{f(\gamma^{-i})}{2^i} \right) + \sum_{K=1}^n \gamma^K \left(\frac{1}{2^{n-K+1}} + \left(\sum_{i=K+1}^n \frac{f(\gamma^{-i})}{2^{i-K+1}} \right) - \frac{f(\gamma^{-K})}{2}\right). \]

Now, let's consider the \SubsetValueStability constraint with $f_{\ell}$ for some parameter $0 < \ell < \infty$. Let's use the fact that $f(\gamma^{-K}) = 1 - \gamma^{2 K\ell}$. Now, we can conclude that:
\begin{align*}
    \ubTV(k, \ell) &\le \left(1 - \sum_{i=1}^n \frac{\gamma^{2 i \cdot \ell}}{2^i} \right) + \sum_{k=1}^n \gamma^{K} \left(-\left(\sum_{i=K+1}^n\frac{\gamma^{2 i \cdot \ell}}{2^{i-K+1}}\right) + \frac{\gamma^{2 K \cdot \ell}}{2} \right) \\
    &\le \left(1 - \sum_{i=1}^n \frac{\gamma^{2 i \cdot \ell}}{2^i} \right) + \frac{1}{2} \left(\sum_{K=1}^n \gamma^{K(2 \ell+1)}\right) - \left(\sum_{K=1}^n \gamma^K\left(\sum_{i=K+1}^n\frac{\gamma^{2 i \cdot \ell}}{2^{i-K+1}}\right)\right).
\end{align*}

The remainder of the analysis boils down to analyzing this expression. Let the RHS of this expression be $E(k, \ell, \gamma)$. We first compute $E(\gamma, \ell) := \lim_{k \rightarrow \infty} E(k, \ell, \gamma)$. Notice that taking a limit as $k \rightarrow \infty$ is the same as taking a limit as $n \rightarrow \infty$. Let's handle each term in the RHS separately.

\begin{enumerate}
    \item For the first term, the limit as $n \rightarrow \infty$ of $\left(1 - \sum_{i=1}^n \frac{\gamma^{2 i\ell}}{2^i} \right)$ is $\left(1 - \sum_{i=1}^{\infty} \frac{\gamma^{2 i\ell}}{2^i} \right) = 1 - \left( \sum_{i=0}^{\infty} \frac{\gamma^{2 i\ell}}{2^i}  - 1 \right) = 2 - \frac{1}{1 - \gamma^{2 \ell}/2}$. 
    \item For the second term, the limit as $n \rightarrow \infty$ is $\frac{1}{2} \left(\sum_{K=1}^n \gamma^{(2 \ell+1)K}\right) = \frac{\gamma^{2 \ell+1}}{2} \left(\sum_{K=0}^{n-1} \gamma^{(2 \ell+1)K}\right)$. Taking a limit as $n \rightarrow \infty$, we obtain $\frac{\gamma^{2 \ell+1}}{2} \frac{1}{1 - \gamma^{2\ell+1}}$.
    \item For the third term, $\frac{1}{2} \left(\sum_{K=1}^n \gamma^K \cdot 2^K \left(\sum_{i=K+1}^n\frac{\gamma^{2 i \cdot \ell}}{2^{i}}\right)\right) = \frac{1}{2} \left(\sum_{K=1}^n \gamma^K \cdot 2^K \frac{\gamma^{2 \ell (K+1)}}{2^{K+1}} \left(\sum_{i=0}^{n-K-1}\frac{\gamma^{2 i \cdot \ell}}{2^{i}}\right)\right)$. We can write this as: $\frac{1}{4 \gamma} \left(\sum_{K=1}^n \gamma^{(K+1)(2 \ell+1)}  \left(\sum_{i=0}^{n-K-1}\frac{\gamma^{2 i \cdot \ell}}{2^{i}}\right)\right)$. Now, we can write this as: 
\[\frac{1}{4 \gamma} \left(\sum_{K=1}^n \gamma^{(K+1)(2 \ell+1)}  \left(\frac{1 - \gamma^{(n-K)l}/2^{n-K}}{1 - \gamma^{2 \ell}/2}\right)\right).\] We can write this as:
\[\frac{1}{4 \gamma} \left[\left(\sum_{K=1}^n \gamma^{(K+1)(2 \ell+1)}  \left(\frac{1}{1 - \gamma^{2 \ell}/2}\right)\right) - \left(\sum_{k=1}^n \gamma^{(K+1)(2 \ell+1)}  \left(\frac{\gamma^{2 \ell (n-K)}/2^{n-K}}{1 - \gamma^{2 \ell}/2}\right)\right) \right].\] 

\begin{enumerate}
    \item The first part is
$\left(\sum_{K=1}^n \gamma^{(K+1)(2 \ell+1)}  \left(\frac{1}{1 - \gamma^{2 \ell}/2}\right)\right) = \left(\frac{1}{1 - \gamma^{2 \ell}/2}\right)\sum_{K=1}^n \gamma^{(K+1)(2 \ell+1)} = \left(\frac{\gamma^{2(2 \ell+1)}}{1 - \gamma^{2 \ell}/2}\right) \sum_{K=0}^{n-1} \gamma^{K(2 \ell+1)}$. Taking a limit as $n \rightarrow \infty$, we obtain $\left(\frac{\gamma^{2(2 \ell+1)}}{1 - \gamma^{2 \ell}/2}\right) \frac{1}{1 - \gamma^{2 \ell+1}}$.
\item For the second part: 
\begin{align*}
    \frac{1}{1 - \gamma^{2 \ell}/2}\left(\sum_{K=1}^n \gamma^{K+1} \gamma^{2(n+1)\ell}/2^{n-K}\right) &= \frac{2}{1 - \gamma^{2 \ell}/2} \frac{\gamma^2 \gamma^{2(n+1)\ell}}{2^n} \left(\sum_{K=1}^n \gamma^{K-1}  2^{K-1}\right) \\
    &= \frac{1}{1 - \gamma^{2 \ell}/2} \frac{\gamma^2 \gamma^{2 (n+1)\ell}}{2^n} \frac{1 - \gamma^n 2^n}{1 - 2 \gamma} \\
    &= \frac{\gamma^{2+2 \ell}}{(1 - \gamma^{2 \ell}/2)(1 - 2 \gamma)} \gamma^{2 n \ell} (1/2^n - \gamma^n).
\end{align*}
This goes to $0$ as $n \rightarrow \infty$. 
\end{enumerate}

Thus, the third term overall becomes $\left(\frac{\gamma^{4\ell+1}}{4(1 - \gamma^{2\ell}/2)}\right) \frac{1}{1 - \gamma^{2\ell+1}}$. 

\end{enumerate}

Putting it all together, we obtain that for every $\gamma < 1$, it holds that:
\[E(\ell, \gamma) = \lim_{k \rightarrow \infty} E(\ell, k, \gamma) = 2 - \frac{1}{1 - \gamma^{2\ell}/2} + \frac{\gamma^{2\ell+1}}{2} \frac{1}{1 - \gamma^{2\ell+1}} -\left(\frac{\gamma^{4\ell+1}}{4(1 - \gamma^{2\ell}/2)}\right) \frac{1}{1 - \gamma^{2\ell+1}}. \]

Now, let $E(\ell) := \lim_{\gamma \rightarrow 1} E(\ell, \gamma)$. We claim that $E(\ell) = \frac{2 \ell}{2 \ell+1}$. Notice that 
\[E(\ell) = \lim_{\gamma \rightarrow 1} \left( 2 - \frac{1}{1 - \gamma^{2\ell}/2} + \frac{\gamma^{2\ell+1}}{2} \frac{1}{1 - \gamma^{2\ell+1}} -\left(\frac{\gamma^{4\ell+1}}{4(1 - \gamma^{2\ell}/2)}\right) \frac{1}{1 - \gamma^{2\ell+1}}\right).\] We claim that $\lim_{\gamma \rightarrow 1} \left( 2 - \frac{1}{1 - \gamma^{2\ell}/2} + \frac{\gamma^{2\ell+1}}{2} \frac{1}{1 - \gamma^{\ell+1}} -\left(\frac{\gamma^{4\ell+1}}{4(1 - \gamma^{2\ell}/2)}\right) \frac{1}{1 - \gamma^{2\ell+1}}\right)  = \frac{2 \ell}{2\ell+1}$.  We see that $2 - \frac{1}{1 - \gamma^{2 \ell}/2}  \rightarrow 0$ as $\gamma \rightarrow 1$. Thus, what remains is: 
\[\frac{\gamma^{2 \ell+1} \left(2(1 - \gamma^{2 \ell}/2) - \gamma^{2 \ell} \right)}{4(1 - \gamma^{2 \ell+1})(1 - \gamma^{2 \ell}/2)}  = \frac{\gamma^{2 \ell+1} \left(2 - 2 \gamma^{2 \ell} \right)}{4(1 - \gamma^{2 \ell+1})(1 - \gamma^{2 \ell}/2)} =  \frac{\gamma^{2 \ell+1}}{2(1 - \gamma^{2 \ell}/2)} \frac{1 - \gamma^{2 \ell}}{1 - \gamma^{2 \ell+1}}.\]
We see that $\frac{\gamma^{2 \ell+1}}{2(1 - \gamma^{2 \ell}/2)} \rightarrow 1$. Notice that $\frac{1 - \gamma^{2 \ell}}{1 - \gamma^{2 \ell+1}} \rightarrow \frac{2 \ell}{2 \ell+1}$, as desired.
Since $\ubTV(k, \ell) \le E(\ell, k, \gamma)$ for every $\gamma$ and every $k$, this means that $\limsup_{k \rightarrow \infty} \ubTV(k, \ell) \le \limsup_{k \rightarrow \infty} E(\ell, k, \gamma) = E(\ell, \gamma)$ for every $\gamma < 1$, and thus $\limsup_{k \rightarrow \infty} \ubTV(k, \ell)\le \lim_{\gamma \rightarrow 1} E(\ell, \gamma) = E(\ell) = \frac{2 \ell }{2 \ell +1}$.  

Now, we analyze $\frac{\lb(\ell)}{\ubTV(k, \ell)}$ in the limit as $k \rightarrow \infty$. Observe that $\frac{\lb(\ell)}{\ubTV(k, \ell)} \ge \frac{1 - \frac{1}{\ell+1} \left(\frac{\ell}{\ell+1}\right)^{\ell}}{E(\ell, k, \gamma)}$. 
In particular, we have that:
\[
\liminf_{k \rightarrow \infty} \frac{\lb(\ell)}{\ubTV(k, \ell)} \ge \liminf_{k \rightarrow \infty} \frac{1 - \frac{1}{\ell+1} \left(\frac{\ell}{\ell+1}\right)^{\ell}}{E(\ell, k, \gamma)} = \frac{1 - \frac{1}{\ell+1} \left(\frac{\ell}{\ell+1}\right)^{\ell}}{\limsup{k \rightarrow \infty}  E(\ell, k, \gamma)}  = \frac{1 - \frac{1}{\ell+1} \left(\frac{\ell}{\ell+1}\right)^{\ell}}{\lim_{k \rightarrow \infty} E(\ell, k, \gamma)} = \frac{1 - \frac{1}{\ell+1} \left(\frac{\ell}{\ell+1}\right)^{\ell}}{E(\ell, \gamma)}.\]
Since this holds for every $\gamma > 1$, we know that:
\begin{align*}
\liminf_{k \rightarrow \infty} \frac{\lb(\ell)}{\ubTV(k, \ell)} &\ge \limsup_{\gamma \rightarrow 1} \frac{1 - \frac{1}{\ell+1} \left(\frac{\ell}{\ell+1}\right)^{\ell}}{E(\ell, \gamma)} \\
&= \frac{1 - \frac{1}{\ell+1} \left(\frac{\ell}{\ell+1}\right)^{\ell}}{\liminf_{\gamma \rightarrow 1} E(\ell, \gamma)} \\
&= \frac{1 - \frac{1}{\ell+1} \left(\frac{\ell}{\ell+1}\right)^{\ell}}{\lim_{\gamma \rightarrow 1} E(\ell, \gamma)} \\
&= \frac{1 - \frac{1}{\ell+1} \left(\frac{\ell}{\ell+1}\right)^{\ell}}{E(\ell)} \\
&= \frac{1 - \frac{1}{\ell+1} \left(\frac{\ell}{\ell+1}\right)^{\ell}}{\frac{2 \ell }{2 \ell+1}} \\
&= \frac{2 \ell+1}{\ell + 1} \frac{\ell + 1 - \left(\frac{\ell}{\ell+1}\right)^{\ell}}{2 \ell} \\
&= \frac{2 \ell+1}{\ell + 1} \left(\frac{1}{2} + \frac{1 - \left(\frac{\ell}{\ell+1}\right)^{\ell}}{2 \ell}\right) \\
&= \frac{2 \ell+1}{\ell + 1} \left(\frac{1}{2} + \frac{1}{2 \ell} \left(1 - \frac{\ell^\ell}{(\ell+1)^\ell} \right) \right). \\
\end{align*}

This can be lower bounded by: $\frac{1}{2 \ell} \left(1 - \frac{\ell^\ell}{(\ell+1)^\ell} \right)$. Now, let's a limit as $\ell \rightarrow 0$ to obtain:
\[\liminf_{\ell \rightarrow 0} \left(\liminf_{k \rightarrow \infty} \frac{\lb(\ell)}{\ubTV(k, \ell)}\right) \ge 
\liminf_{\ell \rightarrow 0} \left( \frac{1}{2 \ell} \left(1 - \frac{\ell^\ell}{(\ell+1)^\ell} \right) \right)  = \infty.\]  
\end{proof}

\end{document}